-------------------------------------------------------------------
\\
\documentstyle[12pt]{article}
\title{{\normalsize{{\hskip 8.5cm} BIHEP-TH-94-9}}\\[-7mm]
{\normalsize{{\hskip 8.5cm} March,~~~~~1994}}\\
$q$-Deformed Chern Class, Chern-Simons\\
 and Cocycle Hierarchy}

\author{  Bo-Yu Hou$^{1)}$, Bo-Yuan Hou$^{2)}$ and
Zhong-Qi Ma$^{3)}$\\
\parbox[t]{15cm}{{\footnotesize {1) Institute of Modern Physics,
Northwest University, Xi'an 710069, P. R. China }}\\[-2mm]
{\footnotesize {2) Graduate School, Chinese Academy of Sciences,
P. O. Box 3908, Beijing 100039, P. R. China}}\\[-2mm]
{\footnotesize {3) Institute of High Energy Physics, P. O. Box 918(4),
Beijing 100039, P. R. of China}}}}

\date{}
\begin{document}
\maketitle

\vspace{20mm}

\begin{abstract}
In this paper, from the $q$-gauge covariant condition we define the
$q$-deformed Killing form and the second $q$-deformed Chern class
for the quantum group $SU_{q}(2)$. Developing Zumino's method
we introduce a $q$-deformed homotopy operator to compute the
$q$-deformed Chern-Simons and the $q$-deformed cocycle hierarchy.
Some recursive relations related to the generalized $q$-deformed
Killing forms are derived to prove the cocycle hierarchy formulas
directly. At last, we construct the $q$-gauge covariant Lagrangian
and derive the $q$-deformed Yang-Mills equation. We find that
the components of the singlet and the adjoint representation are
separated in the $q$-deformed Chern class, $q$-deformed cocycle
hierarchy and the $q$-deformed Lagrangian,
although they are mixed in the commutative relations of BRST algebra.

\end{abstract}

\newpage
\noindent
{\bf 1. INTRODUCTION}

Recently, quantum groups have attracted increasing attention.
Manin$^{[1]}$ suggested a general construction of quantum groups as
linear transformations upon the quantum superplane.
Woronowicz$^{[2]}$ elaborated the connection of quantum groups with
noncommutative differential geometry. The differential
calculus on the quantum hyperplane was developed by Wess and
Zumino $^{[3]}$. There have been a lot of papers treating the
differential calculus on quantum groups and the deformed gauge
theories from various viewpoints$^{[4-9]}$.

Aschieri and Castellani$^{[10]}$ gave a pedagogical introduction to
the differential calculus on quantum groups by stressing at all
stages the connection with the classical case ($q~ \rightarrow ~1$).
In the paper [11] on deformed gauge theories, Castellani tried to
construct the $q$-deformed Lagrangian and the $q$-deformed Killing
form. In his paper and also in the present paper, the spacetime is taken
to be the ordinary commutative Minkowski spacetime, while the
$q$-structure resides on the fiber, the gauge potentials being
non-commutating.  In our opinion, his $q$-deformed
Killing form is not very satisfactory due to the unclear parameter.

Watamura$^{[12]}$ carefully discussed the $q$-deformed gauge
transformation and $q$-deformed BRST formalism. He extracted appropriate
properties from the non-deformed BRST formalism and imposed them as
the condition which new algebra should satisfy so that the
$q$-deformed BRST formalism is the $q$-analogue of the non-deformed
one. However, he did not find the $q$-deformed Killing form
that is the key for constructing the $q$-deformed Chern class
and the $q$-deformed Lagrangian.

{}From the $q$-gauge covariant condition we define the $q$-deformed
Killing form and the second $q$-deformed Chern class for the
quantum group $SU_{q}(2)$ under the BRST formalism constructed by
Watamura. Developing Zumino's method$^{[13]}$,
we introduce a $q$-deformed homotopy operator to compute the
$q$-deformed Chern-Simons and the $q$-deformed cocycle hierarchy.
At last, we construct the $q$-gauge covariant Lagrangian
and derive the $q$-deformed Yang-Mills equation. We find that
the components of the singlet and the adjoint representation are
separated in the $q$-deformed Chern class, $q$-deformed cocycle
hierarchy and the $q$-deformed Lagrangian, although they are mixed
in the commutative relations of BRST algebra. The formalism
discussed in this paper can be generalized to the quantum groups
$SU_{q}(N)$ straightforwardly.

The plan of this paper is as follows. In Sec. 2 we rewrite some
formulas of noncommutative differential geometry on quantum groups
$SU_{q}(2)$ based on the right-invariant quantities that appear in
BRST formalism,
although the similar ones related to the left-invariant quantities
were explicitly given in [10]. We have to unify the notations used
in Refs. [10] and [12], and make the formalism more accessible.
The explicit values of some quantities are listed in Appendix.
The $q$-deformed Killing form $g_{IJ}$ is defined in Sec. 3 by
requiring the invariance of the second $q$-deformed Chern class.
We also derive some important recursive relations related to the
generalized $q$-deformed Killing forms. In Sec. 4 we
introduce a $q$-deformed homotopy operator, that is the analogue of
Zumino's homotopy operator$^{[13]}$, to compute the $q$-deformed Chern-Simons,
from that we calculate the $q$-deformed cocycle hierarchy in Sec. 5.
We directly prove the cocycle hierarchy formulas by the recursive relations
given in Sec. 3. At last, we construct the $q$-deformed Lagrangian
and derive the $q$-deformed Yang-Mills equation in Sec. 6.

\vspace{20mm}
\noindent
{\bf 2. Non-Commutative Differential Geometry and BRST Formalism}

The solution of the simple Yang-Baxter equation$^{[14]}$,
related to the fundamental representation of
$SU_{q}(2)$, is well known$^{[15]}$ as follows:
$$\breve{R}_{q}~=~{\cal P}_{S} ~-~q^{2}~{\cal P}_{A},~~~
\breve{R}_{q}^{-1}~=~{\cal P}_{S} ~-~q^{-2}~{\cal P}_{A}
\eqno (2.1)$$

\noindent
where ${\cal P}$ denotes the projection operator:
$${\cal P}_{S}~=~{ 1 \over [2]}~~\left(\begin{array}{cccc} [2]&0&0&0\\
0&q&1&0\\ 0&1&q^{-1}&0\\ 0&0&0&[2] \end{array} \right) ~~~~
{\cal P}_{A}~=~{ 1 \over [2]}~~\left(\begin{array}{cccc} 0&0&0&0\\
0&q^{-1}&-1&0\\ 0&-1&q&0\\ 0&0&0&0 \end{array} \right) \eqno (2.2) $$

\noindent
and the $q$-number is defined as usual:
$$[m]~=~{\displaystyle q^{m}~-~q^{-m} \over \displaystyle q~-~q^{-1}}
\eqno (2.3)$$

A quantum group is introduced as the non-commutative Hopf algebra
${\cal A}=Fun(G)$ obtained by continuous deformations of the Hopf algebra
of the function of a Lie group. The associative algebra ${\cal A}$ is
freely generated by non-commutating matrix entries $T^{a}_{~b}$
satisfying the relation:
$$\begin{array}{c}
\breve{R}_{q}~(T~\otimes~T) ~=~(T~\otimes~T)~\breve{R}_{q},~~~~
T^{a}_{~b}~=~\left(\begin{array}{cc} \alpha&\beta \\ \gamma &\delta
\end{array} \right) \\
\breve{R}^{ab}_{~ef}~T^{e}_{~c}~T^{f}_{~d}~=~
T^{a}_{~e}~T^{b}_{~f}~\breve{R}^{ef}_{~cd} \end{array}
 \eqno (2.4) $$
$$\begin{array}{c}
\alpha \beta ~=~q~\beta \alpha ,~~~~~\alpha \gamma ~=~q~\gamma \alpha \\
\beta \delta ~=~q~\delta \beta ,~~~~ \gamma \delta ~=~q~\delta \gamma,
{}~~~~\beta \gamma ~=~\gamma \beta, \\
\alpha \delta ~-~q~ \beta \gamma~=~\delta \alpha ~-~q^{-1}~\beta
\gamma ~=~ \det T ~=~1 \end{array} \eqno (2.5) $$

\noindent
where the repeated indices, for example $a$, are summed over 1 and $2$.
Throughout this paper, if without special notification, summation of
the repeated indices is understood.

As a Hopf algebra, the matrix elements $T^{a}_{~b}$ satisfy the
Hopf algebraic relations:
$$\begin{array}{c}
\Delta(T^{a}_{~b})~=~T^{a}_{~d}~\otimes~T^{d}_{~b},~~~
\epsilon(T^{a}_{~b})~=~\delta^{a}_{b} \\
\kappa(T^{a}_{~b})~=~(T^{-1})^{a}_{~b} ,~~~~
\kappa^{2}(T^{a}_{~b})~=~D^{a}_{~c}~T^{c}_{~d}~(D^{-1})^{d}_{~b}
\end{array} \eqno (2.6) $$
$$(T^{-1})^{i}_{~j}~=~\left(\begin{array}{cc} \delta&-q^{-1}\beta \\
-q \gamma &\alpha \end{array} \right) $$

\noindent
where $D$ is a diagonal matrix:
$$D^{1}_{~1}~=~q^{-1},~~~~
D^{2}_{~2}~=~q \eqno (2.7) $$

Introduce $q$-Pauli matrices:
$$\begin{array}{c}
q[2]^{1/2}~(\sigma^{0})_{ab}~=~q^{-1}[2]^{1/2}~(\sigma_{0})^{ab}
{}~=~\epsilon^{ab}~=~-~\epsilon_{ab}
{}~=~\left(\begin{array}{cc} 0&-q^{-1/2}
 \\ q^{1/2}&0 \end{array} \right) \\[1mm]
(\sigma^{3})_{ab}~=~(\sigma_{3})^{ab}~=~-~[2]^{-1/2}~
\left(\begin{array}{cc} 0&q^{1/2} \\ q^{-1/2} &0  \end{array} \right) \\
(\sigma^{+})_{ab}~=~(\sigma_{+})^{ab}~=~
\left(\begin{array}{cc} 0&0 \\ 0 &-1  \end{array} \right) ,~~~~~
(\sigma^{-})_{ab}~=~(\sigma_{-})^{ab}~=~
\left(\begin{array}{cc} 1&0 \\ 0 &0  \end{array} \right) \\
(\sigma^{I})^{b}_{a}~=~(\sigma^{I})_{ad}~\epsilon^{db} ,
{}~~~~(\sigma_{I})^{b}_{a}~=~(\sigma_{I})^{bd}~\epsilon_{ad}
 \end{array} \eqno (2.8) $$
$$\begin{array}{ll}
(\sigma^{I})_{ab}~(\sigma_{J})^{ab}~=~\delta^{I}_{J},~~~~&
(\sigma^{I})_{ab}~(\sigma_{I})^{cd}~=~\delta^{c}_{a}~\delta^{d}_{b}\\
(\sigma^{I})_{a}^{b}~(\sigma_{J})_{b}^{a}~=~\delta^{I}_{J},~~~~&
(\sigma^{I})^{a}_{c}~(\sigma_{I})^{d}_{b}~
=~\delta^{a}_{b}~\delta^{d}_{c} \\
({\cal P}_{A})^{ab}_{~cd}~=~(\sigma_{0})^{ab}~(\sigma^{0})_{cd},
{}~~~~&({\cal P}_{S})^{ab}_{~cd}~=~(\sigma_{i})^{ab}~(\sigma^{i})_{bd}
\end{array} \eqno (2.9) $$

\noindent
Throughout this paper, the capital latin, like $I$, runs over $0$,
$+$, $3$, and $-$, the small latin, like $i$, runs over $+$, $3$,
and $-$, but the first a few small latin, like $a$, runs over $1$ and $2$.

The direct product representation $M^{~a_{2}b_{1}}_{a_{1}~~~b_{2}}$ of the
quantum group $SU_{q}(2)$ is defined by:
$$M^{~a_{2}b_{1}}_{a_{1}~~~b_{2}}~=~
\kappa^{2}(T^{a_{2}}_{~b_{2}})~\kappa(T^{b_{1}}_{~a_{1}})
{}~=~D^{a_{2}}_{d_{2}}~T^{d_{2}}_{~d_{1}}(D^{-1})^{d_{1}}_{b_{2}}
{}~\kappa(T^{b_{1}}_{~a_{1}}) \eqno (2.10) $$

\noindent
Now, the singlet and the adjoint representation can be separated
from $M^{~a_{2}b_{1}}_{a_{1}~~~b_{2}}$:
$$\begin{array}{c}
M^{I}_{~J}~=~(\sigma^{I})^{a_{1}}_{a_{2}}~(\sigma_{J})^{b_{2}}_{b_{1}}
{}~M^{~a_{2}b_{1}}_{a_{1}~~~b_{2}} \\
M^{i}_{~0}~=~M^{0}_{~i}~=~0,~~~~M^{0}_{~0}~=~{\bf 1 } \\
\kappa^{2}(M^{I}_{~J})~=~D^{I}_{K}~M^{K}_{~L}~(D^{-1})^{L}_{~J}\\
D_{~0}^{0}~=~D_{~3}^{3}~=~1,~~~~D_{~+}^{+}~=~q^{2},~~~~D_{~-}^{-}~=~q^{-2}
\end{array} \eqno (2.11) $$

\noindent
where $D$ is a diagonal matrix.

The linear functionals$^{[16]}$, $(L^{\pm})^{a}_{~b}$, defined by their
values on the entries $T^{a}_{~b}$, belong to the dual Hopf algebra
${\cal A}'$:
$$(L^{+})^{a}_{~b}\left(T^{c}_{~d}\right)~=~q^{1/2}~
(\breve{R}^{-1}_{q})^{ac}_{~~db}~,~~~~
(L^{-})^{a}_{~b}\left(T^{c}_{~d}\right)~=~q^{-1/2}~
(\breve{R}_{q})^{ac}_{~~db} \eqno (2.12) $$
$$\begin{array}{c}
\Delta'\left( (L^{\pm})^{a}_{~b} \right)~=~(L^{\pm})^{a}_{~c}
{}~\otimes~(L^{\pm})^{c}_{~b},~~~~
\epsilon'\left( (L^{\pm})^{a}_{~b}\right)~=~\delta^{a}_{b} \\
\kappa'\left( (L^{\pm})^{a}_{~c}\right)~\left(L^{\pm}\right)^{c}_{~b}
{}~=~\delta^{a}_{b}~=~
(L^{\pm})^{a}_{~c}~\kappa'\left((L^{\pm})^{c}_{~b}\right)
\end{array} \eqno (2.13) $$
$$\breve{R}_{q}^{21}~L_{1}^{\pm}~L_{2}^{\pm}~=~
L_{1}^{\pm}~L_{2}^{\pm}~\breve{R}_{q}^{21} ,~~~~
\breve{R}_{q}^{21}~L_{1}^{-}~L_{2}^{+}~=~
L_{1}^{+}~L_{2}^{-}~\breve{R}_{q}^{21} \eqno (2.14) $$

Now, we are going to construct the bimodule $\Gamma$. Usually,
$\Gamma$ is called the space of quantum one-form $^{[10]}$, but in
the present paper we call it the space with ghost number 1 in order
to avoid confusion with the spacetime one form space. And following
Watamura$^{[12]}$, we call the first-order differential calculus on
${\cal A}$ the BRST transformation operator, denoted by $\delta$:
$$\begin{array}{l}
\delta: ~{\cal A}~\rightarrow~\Gamma \\
\rho~=~a_{K}~\delta~ b_{K}~\in~\Gamma,~~~~{\rm if}~~\rho \in \Gamma
\end{array} \eqno (2.15) $$

\noindent
where $ a_{K}, ~b_{K} ~\in ~{\cal A}$.

A left action $\Delta_{L}$ and a right action $\Delta_{R}$ of the
quantum group on $\Gamma$ are defined as follows:
$$\begin{array}{l}
\Delta_{L}(a\delta b)~=~\Delta(a)~( id~ \otimes ~\delta)~\Delta(b),
{}~~~\Delta_{L}~:~\Gamma~\rightarrow~{\cal A}~\otimes ~\Gamma \\
\Delta_{R}(a\delta b)~=~\Delta(a)~( \delta ~\otimes ~id)~\Delta(b),
{}~~~\Delta_{R}~:~\Gamma~\rightarrow~\Gamma ~\otimes ~{\cal A}
\end{array} \eqno (2.16) $$

\noindent
For the tensor product between elements $\rho$, $\rho'\in \Gamma$,
$$\begin{array}{c}
(\rho~a)~\otimes~\rho'~=~\rho~\otimes~(a~\rho')\\
a(\rho~\otimes~\rho')~=~(a~\rho)~\otimes~\rho',~~~~
(\rho~\otimes~\rho')a~=~\rho~\otimes~(\rho'~a) \end{array} \eqno (2.17) $$

\noindent
the left and right actions on $\Gamma \otimes \Gamma$ are defined by:
$$\begin{array}{l}
\Delta_{L}~:~\Gamma~\otimes~\Gamma~\rightarrow~{\cal A}~\otimes~
\Gamma~\otimes~\Gamma \\
\Delta_{R}~:~\Gamma~\otimes~\Gamma~\rightarrow~\Gamma~
\otimes~\Gamma~\otimes ~{\cal A} \end{array} \eqno (2.18) $$

\noindent
For example,
$$\begin{array}{l}
\Delta_{L}(\rho_{1})~=~a_{1}~\otimes~\rho_{1}',~~~
\Delta_{L}(\rho_{2})~=~a_{2}~\otimes~\rho_{2}' \\
\Delta_{L}(\rho_{1}~\otimes ~\rho_{2})~=~a_{1}~a_{2}~\otimes~\rho_{1}'
{}~\otimes~\rho_{2}' \end{array}\eqno (2.19) $$

\noindent
{}From the definition we have:
$$\begin{array}{rl}
(\epsilon~ \otimes ~id)~\Delta_{L}(\rho)~=~\rho,~&~
(id~ \otimes ~\epsilon)~\Delta_{R}(\rho)~=~\rho \\
(\Delta ~\otimes ~id)~\Delta_{L}~=~(id ~\otimes~ \Delta_{L})~\Delta_{L},~&~
(id ~\otimes~ \Delta)~\Delta_{R}~=~(\Delta_{R} ~\otimes~ id)~\Delta_{R} \\
(id~ \otimes~ \Delta_{R} )~\Delta_{L}&=~(\Delta_{L}~ \otimes~ id)
{}~\Delta_{R}$$
\end{array} \eqno (2.20) $$
The bases of the right-invariant elements of $\Gamma$ are denoted by
$\eta^{J}$, satisfying:
$$\begin{array}{l}
\Delta_{R}(\eta^{J})~=~\eta^{J} ~\otimes ~{\bf 1} ,~~~~
\Delta_{L}(\eta^{J})~=~M^{J}_{~K} ~\otimes ~\eta^{K} \\
a ~\eta^{J}~=~\eta^{K}(a~ *~ L^{J}_{K}),~~~a\in {\cal A},~~~
L^{J}_{K}\in {\cal A}'  \end{array} \eqno (2.21) $$

\noindent
where
$$\begin{array}{l}
L^{I}_{J}~=~(\kappa')^{-1}\left((L^{-})^{a_{2}}_{b_{2}}\right)
{}~(L^{+})^{b_{1}}_{a_{1}}~
(\sigma^{I})^{a_{1}}_{a_{2}}~(\sigma_{J})^{b_{2}}_{b_{1}}\\
L^{I}_{J}(ab)~=~L^{K}_{J}(a)~L^{I}_{K}(b),~~~~
L^{I}_{J}({\bf 1})~=~\delta^{I}_{J} \end{array} \eqno (2.22) $$
$$\begin{array}{l}
(\rho ~*~ L^{I}_{J})~=~(L^{I}_{J}~ \otimes ~id)~\Delta_{L}(\rho) \\
(L^{I}_{J}~ *~ \rho)~=~(id ~\otimes ~L^{I}_{J})~\Delta_{R}(\rho)
\end{array} \eqno (2.23) $$

\noindent
Note that:
$$\Delta_{L}(\eta^{0})~=~{\bf 1}~ \otimes ~\eta^{0} \eqno (2.24) $$

Now, the BRST transformation on ${\cal A}$ can be computed
as follows:
$$\delta a~=~\displaystyle { ig \over \lambda} ~(\eta^{0} ~a ~- ~a ~\eta^{0})
{}~=~\eta^{J} ~(a~ * ~\chi_{J}) \eqno (2.25) $$

\noindent
where
$$\begin{array}{c}
\lambda = q-q^{-1},~~~~\lambda^{2}~+~4~=~[2]^{2},~~~~
\lambda^{2}~+~3~=~[3] \\
\lambda^{2}~+~2~=~\displaystyle {[4]\over[2]},
{}~~~~\lambda^{2}~+~1~=~\displaystyle {[6]\over[3][2]}
\end{array} \eqno (2.26) $$

\noindent
and
$$\begin{array}{l}
\chi_{J}~=~\displaystyle { ig \over \lambda}~\left(~ \epsilon~
\delta^{0}_{J}~-~L^{0}_{J}~\right) \\
\chi_{J}(ab)~=~\chi_{J}(a)~\epsilon(b)~+~L^{K}_{J}(a)~\chi_{K}(b),
{}~~~~\chi_{J}({\bf 1})~=~0 \\
\chi_{J}~\chi_{K}~=~(\chi_{J}~\otimes~\chi_{K})~\Delta
\end{array} \eqno (2.27) $$

\noindent
where $\chi_{J} \in {\cal A}'$ are the $q$-analogues of the tangent vectors
at the identity element of the group, and $~*\chi_{J}$ are the
analogues of right invariant vector fields$^{[10]}$. The explicit
values of $L^{J}_{K}$ and $\chi_{J}$ on ${\cal A}$ are given in Appendix.

The basis of the left-invariant element of $\Gamma$ is easy to be
calculated from $\eta^{J}$:
$$\omega^{J}~=~\eta^{K}~\kappa^{-1}(M^{J}_{~K}),~~~~\Delta_{L}(\omega^{J})
{}~=~{\bf 1} ~\otimes ~\omega^{J},~~~~\Delta_{R}(\omega^{J})~=~
\omega^{K} ~\otimes ~\kappa^{-1}(M^{J}_{~K}) \eqno (2.28) $$

As the analogue of the ordinary permutation operator, a bimodule
automorphism $\Lambda$ in $\Gamma \otimes \Gamma$ is defined by:
$$\begin{array}{c}
\Lambda(\omega^{J}~\otimes ~\eta^{K})~=~
\eta^{K}~\otimes ~\omega^{J} \\
\Lambda(a~\tau)~=~a~\Lambda(\tau),~~~~
\Lambda(\tau a)~=~\Lambda(\tau)~a, ~~~~ a\in {\cal A},~~\tau\in
\Gamma\otimes \Gamma \end{array}  \eqno (2.29) $$

\noindent
Thus, we have
$$\Lambda(\eta^{I}~\otimes ~\eta^{J})~=~\Lambda^{IJ}_{~~KL}
{}~\eta^{K}~\otimes ~\eta^{L},~~~~\Lambda^{IJ}_{~~KL}~=~L^{J}_{K}
(M^{I}_{~L}) \eqno (2.30) $$

\noindent
The non-vanishing components of $\Lambda^{IJ}_{~~KL}$ are listed as follows:
$$\begin{array}{c}
\Lambda^{ij}_{~~k\ell}~=~(\Lambda^{-1})^{ij}_{~~k\ell}
{}~=~\delta^{i}_{k}~\delta^{j}_{\ell}~-~f_{ij}^{n}~f_{k\ell}^{n},~~~
\Lambda^{00}_{~~00}~=~(\Lambda^{-1})^{00}_{~~00}~=~1 \\
\Lambda^{i0}_{~~jk}~=~(\Lambda^{-1})^{0i}_{~~jk}~=~\lambda~f_{jk}^{i},~~~
\Lambda^{jk}_{~~0i}~=~(\Lambda^{-1})^{jk}_{~~i0}~=~-~\lambda~f_{jk}^{i} \\
\Lambda^{0i}_{~~j0}~=~(\Lambda^{-1})^{i0}_{~~0j}~=~\delta^{i}_{j} ,~~~
\Lambda^{i0}_{~~0j}~=~(\Lambda^{-1})^{0i}_{~~j0}~=~
(\lambda^{2}+1)~\delta^{i}_{j}  \end{array} \eqno (2.31) $$

\noindent
where the non-vanishing components of $f_{bd}^{a}$ are:
$$\begin{array}{c}
f_{+3}^{+}~=~f_{3-}^{-}~=~q,~~~f_{3+}^{+}~=~f_{-3}^{-}~=~-~q^{-1}\\
f_{-+}^{3}~=~-~f_{+-}^{3}~=~1,~~~f_{33}^{3}~=~\lambda
\end{array} \eqno (2.32) $$

For given tensors $X^{ab}_{~~cd}$ and $Y^{ab}_{~~cd}$,
define a operator$^{[12]}$:
$$\begin{array}{rl}
(X,Y)^{IJ}_{~~KL}&=~V^{I}_{~I'}~V^{J}_{~J'}~
(\sigma^{I'})_{a_{1}a_{2}}~(\sigma^{J'})_{b_{1}b_{2}}~
X^{a_{1}e_{1}}_{~~c_{1}e_{3}}
\left(\breve{R}_{q}\right)^{a_{2}b_{1}}_{~~e_{1}e_{2}}\\
&~~\cdot~Y^{e_{2}b_{2}}_{~~e_{4}d_{2}}
\left(\breve{R}_{q}^{-1}\right)^{e_{3}e_{4}}_{~~c_{2}d_{1}}
(\sigma_{K'})^{c_{1}c_{2}}~(\sigma_{L'})^{d_{1}d_{2}}~(V^{-1})^{K'}_{~K}~
(V^{-1})^{L'}_{~L} \end{array} \eqno (2.33) $$

\noindent
where $V$ is a diagonal matrix:
$$V^{0}_{~0}~=~-~q^{2},~~~V^{+}_{~+}~=~V^{3}_{~3}~=~V^{-}_{~-}~=~1$$

\noindent
Hence, we obtain four projection operators
$\left({\cal P}_{r},{\cal P}_{t}\right)$, $r$, $t=S$ or $A$,
satisfying:
$$\begin{array}{c}
\left({\cal P}_{r},{\cal P}_{t}\right)~
 \left({\cal P}_{r'},{\cal P}_{t'}\right)~=~
\delta_{rr'}~\delta_{tt'}~\left({\cal P}_{r},{\cal P}_{t}\right)
,~~~~r,~t ~=~S~~{\rm or}~~A\\
{\cal P}_{Si}~\equiv~\left({\cal P}_{S},{\cal P}_{S}\right)~+~
\left({\cal P}_{A},{\cal P}_{A}\right) ,~~~~
{\cal P}_{Ad}~\equiv~\left({\cal P}_{A},{\cal P}_{S}\right)~+~
\left({\cal P}_{S},{\cal P}_{A}\right)\\
{\cal P}_{Si}~+~{\cal P}_{Ad}~=~{\bf 1} \end{array} \eqno (2.34) $$

\noindent
Operators ${\cal P}_{Si}$ and ${\cal P}_{Ad}$ are called the
projection operators for the singlet and adjoint representation,
respectively.

{}From direct check we find:
$$\begin{array}{l}
\Lambda^{IJ}_{~~KL}~=~\left(\breve{R},\breve{R}^{-1}\right)^{IJ}_{~~KL}\\
{}~~~=~\left({\cal P}_{S},{\cal P}_{S}\right)^{IJ}_{~~KL}~-~q^{-2}~
\left({\cal P}_{S},{\cal P}_{A}\right)^{IJ}_{~~KL}~-~q^{2}~
\left({\cal P}_{A},{\cal P}_{S}\right)^{IJ}_{~~KL}~+~
\left({\cal P}_{A},{\cal P}_{A}\right)^{IJ}_{~~KL} \\
(\Lambda^{-1})^{IJ}_{~~KL}
{}~=~\left(\breve{R}^{-1},\breve{R}\right)^{IJ}_{~~KL}\\
{}~~~=~\left({\cal P}_{S},{\cal P}_{S}\right)^{IJ}_{~~KL}~-~q^{2}~
\left({\cal P}_{S},{\cal P}_{A}\right)^{IJ}_{~~KL}~-~q^{-2}~
\left({\cal P}_{A},{\cal P}_{S}\right)^{IJ}_{~~KL}~+~
\left({\cal P}_{A},{\cal P}_{A}\right)^{IJ}_{~~KL}
\end{array} \eqno (2.35) $$

\noindent
Therefore, $\Lambda^{IJ}_{~~KL}$ satisfy the Yang-Baxter equation:
$$\Lambda^{IJ}_{~~LM}~\Lambda^{MK}_{~~NR}~\Lambda^{LN}_{~~PQ}~=~
\Lambda^{JK}_{~~LM}~\Lambda^{IL}_{~~PN}~\Lambda^{NM}_{~~QR}
\eqno (2.36) $$

\noindent
Since the eigenvalues of $\breve{R}_{q}$ are $1$ and $-q^{2}$,
the eigenvalues of $\Lambda$ matrix are $1$, $-q^{2}$ and $-q^{-2}$:
$$\left(\Lambda~+~q^{2}\right)~\left(\Lambda~+~q^{-2}\right)
{}~\left(\Lambda~-~{\bf 1}\right)~=~0 \eqno (2.37) $$

Now, defining the exterior product of the elements in $\Gamma$:
$$\rho~\wedge \rho'~\equiv~\rho~\otimes \rho'~-~\Lambda\left(
\rho~\otimes \rho'\right) \eqno (2.38) $$

\noindent
we have:
$$\eta^{I}~\wedge~\eta^{J}~=~\left(\delta^{I}_{K}~\delta^{J}_{L}
{}~-~\Lambda^{IJ}_{~~KL}\right)~\left(\eta^{K}~\otimes~\eta^{L}\right) $$

\noindent
{}From Eqs.(2.35) and (2.37) we know that $\eta^{I}~\wedge~\eta^{J}$
is annihilated by the projection operator ${\cal P}_{Si}$:
$$\begin{array}{rl}
{\cal P}_{Si}&=~[2]^{-2}~\left(\Lambda~+~q^{2}\right)~
\left(\Lambda~+~q^{-2}\right) \\
&=~[2]^{-2}~\left\{\Lambda~+~\Lambda^{-1}~+~(\lambda^{2}+2)~{\bf 1}
\right\} \end{array} \eqno (2.39) $$

\noindent
namely,
$$({\cal P}_{Si})^{IJ}_{~~KL}~\left(\eta^{K}~\wedge~\eta^{L}\right)~=~0
 \eqno (2.40) $$

\noindent
The explicit relations are given in Appendix.

The projection operator ${\cal P}_{Ad}$ now can be expressed
as follows:
$${\cal P}_{Ad}~=~[2]^{-2}~\left\{{\bf 2} ~-~\Lambda~-~\Lambda^{-1}\right\}
\eqno (2.41) $$

\noindent
The values of ${\cal P}_{Si}$ and ${\cal P}_{Ad}$ are listed in (A.4).
It is interesting to notice that we can introduce a projection
operator $\hat{{\cal P}}_{Ad}$ with only the adjoint components:
$$\begin{array}{c}
\left(\hat{{\cal P}}_{Ad}\right)^{ij}_{~~k\ell}~=~\displaystyle {[2]^{2}
\over 2(\lambda^{2}+2)}~\left({\cal P}_{Ad}\right)^{ij}_{~~k\ell}
{}~=~(\lambda^{2}+2)^{-1}~f^{t}_{ij}~f^{t}_{k\ell}
\end{array} \eqno (2.42) $$
$$\begin{array}{l}
\left(\hat{{\cal P}}_{Ad}\right)^{ij}_{~~rs}~
\left(\hat{{\cal P}}_{Ad}\right)^{rs}_{~~k\ell}
{}~=~\left(\hat{{\cal P}}_{Ad}\right)^{ij}_{~~k\ell} \\
\left(\hat{{\cal P}}_{Ad}\right)^{ij}_{~~k\ell}~\left(\eta^{k}~
\wedge~ \eta^{\ell} \right)~=~\eta^{i} ~\wedge ~\eta^{j} \\
\left(\hat{{\cal P}}_{Ad}\right)^{ij}_{~~k\ell}~{\bf C}_{ij}^{~~t}
{}~=~{\bf C}_{k\ell}^{~~t}
\end{array} \eqno (2.43) $$

Now, generalize the BRST transformation to $\Gamma^{\otimes n}$.
{}From (2.25) we obtain:
$$\delta \eta^{I}~=~\eta^{J} \otimes \left( \eta^{I} * \chi_{J}\right)
{}~=~{\bf C}_{JK}^{~~I}~\eta^{J} \otimes \eta^{K},~~~~
{\bf C}_{JK}^{~~I}~=~\chi_{J}\left(M^{I}_{~K}\right)
\eqno (2.44) $$

\noindent
It is called the $q$-deformed Cartan-Maurer equation.
${\bf C}_{JK}^{~~I}$ are the $q$-deformed structure constants. The
explicit values of ${\bf C}_{IJ}^{~~K}$ are:
$${\bf C}_{JK}^{~~0}~=~{\bf C}_{J0}^{~~K}~=~0,~~~~{\bf C}_{0k}^{~~j}~=~
-~ig\lambda~\delta_{k}^{j},~~~~{\bf C}_{jk}^{~~i}~=~-~ig~f_{jk}^{i}
 \eqno (2.45) $$

\noindent
where $f_{jk}^{i}$ were given in (2.32). Now, from (2.31) we have:
$$\begin{array}{rl}
\delta \eta^{0}&=~0 \\
\delta \eta^{i}&=~{\bf C}_{JK}^{~~i}~\eta^{J}~\otimes~\eta^{K} \\
&=~-~ig\lambda~\eta^{0}\otimes \eta^{i}
{}~-~ig~f_{jk}^{i}~\eta^{j}\otimes \eta^{k}\\
&=~\{ig/\lambda\}~\left\{\eta^{0}\otimes \eta^{i}
{}~+~\eta^{i}\otimes \eta^{0}~-~\left(\Lambda^{0i}_{~~JK}~+~
\Lambda^{i0}_{~~JK}\right)~\eta^{J}\otimes \eta^{K} \right\}\\
&=~\{ig/\lambda\}~\left\{\eta^{0}\wedge \eta^{i}
{}~+~\eta^{i}\wedge \eta^{0} \right\} \end{array} $$

\noindent
namely,
$$\delta \eta^{I}~=~\displaystyle {ig \over \lambda}~
\left\{\eta^{0}\wedge \eta^{I}
{}~+~\eta^{I}\wedge \eta^{0} \right\}
{}~=~(\lambda^{2}+2)^{-1}~{\bf C}_{jk}^{~~I}~
\left(\eta^{j}\wedge \eta^{k}\right)
\eqno (2.46) $$

Generally, we define:
$$\begin{array}{l}
\delta: ~\Gamma^{\otimes n}~\rightarrow~\Gamma^{\otimes n} ~\wedge~\Gamma \\
\delta \rho~=~\displaystyle { ig \over \lambda} ~\left\{\eta^{0}
{}~\wedge~\rho ~-~(-1)^{n} ~\rho ~\wedge~ ~\eta^{0} \right\},
{}~~~\rho\in \Gamma^{\otimes n} \end{array} \eqno (2.47)  $$

{}From the condition $\delta^{2}a=0$, the functionals $\chi_{J}$ span the
"$q$-deformed Lie algebra":
$$\chi_{I}~\chi_{J}~-~\Lambda^{KL}_{~~IJ}~\chi_{K}~\chi_{L}~=~
{\bf C}_{IJ}^{~~K}~\chi_{K} \eqno (2.48) $$

\noindent
Thus, the $q$-deformed structure constants
satisfy the  $q$-deformed Jacobi identities:
$${\bf C}_{IR}^{~~P}{\bf C}_{JS}^{~~R}~-~\Lambda^{KL}_{~~IJ}~
{\bf C}_{KR}^{~~P}{\bf C}_{LS}^{~~R}~=~{\bf C}_{IJ}^{~~R}
{\bf C}_{RS}^{~~P}\eqno (2.49) $$

\noindent
For the adjoint components we obtain from (2.31):
$$\left(\hat{{\cal P}}_{Ad}\right)^{k\ell}_{~~ij}~
{\bf C}_{kr}^{~~p}{\bf C}_{\ell s}^{~~r}
{}~=~\displaystyle {\lambda^{2}+1 \over \lambda^{2}+2 }~{\bf C}_{ij}^{~~r}
{\bf C}_{rs}^{~~p}\eqno (2.50) $$

Now, we are in the position to introduce gauge potentials and
BRST algebra ${\cal B}$. There are two operators in ${\cal B}$:
the spacetime exterior derivative operator $d$ and the BRST
transformation operator $\delta$. Watamura$^{[12]}$ defined $\eta^{J}$
in the bimodule $\Gamma$ to be the ghost field in the BRST algebra,
that has the ghost number 1, but the degree of form 0. The gauge
potential $A^{J}$ has the degree of form 1, but the ghost number 0.
The operator $\delta$ increases the ghost number by one, and the
operator $d$ increases the degree of form by one. Neglecting the
matter field, we only discuss four fields in the BRST algebra
${\cal B}$: $\eta^{J}$, $d\eta^{J}$, $A^{J}$, and $dA^{J}$, which
satisfy the following algebraic relations.

Firstly, we introduce an index $n$ that is equal to the difference between
the degree of form and the ghost number. The indices $n$ for $\eta^{J}$,
$d\eta^{J}$, $A^{J}$ and $dA^{J}$ are -1, 0, 1, and 2, respectively.
Both $\delta$ and $d$ operations satisfy the Leibniz rule in the
graded sense for the index $n$, and are nilpotent operators:
$$\begin{array}{l}
\delta^{2}~=~0,~~~~d^{2}~=~0,~~~~d~\delta~+~\delta~d~=~0 \\
d(XY)~=~(dX)Y~+~(-1)^{n_{x}}~X(dY) \\
\delta(XY)~=~(\delta X)Y~+~(-1)^{n_{x}}~X(\delta Y)
\end{array} \eqno (2.51) $$

\noindent
where $n_{x}$ is the index of $X$. Both $d$ and $\delta$ are covariant
for the left and right actions: For any element
$\rho \in {\cal B}$ they satisfy:
$$\begin{array}{l}
\Delta_{L}(\delta \rho)~=~(id~\otimes ~\delta)\Delta_{L}(\rho),~~~
\Delta_{L}(d\rho)~=~(id~\otimes ~d)\Delta_{L}(\rho) \\
\Delta_{R}(\delta \rho)~=~(\delta~\otimes ~id)\Delta_{R}(\rho),~~~
\Delta_{R}(d\rho)~=~(d~\otimes ~id)\Delta_{R}(\rho) \end{array}
\eqno (2.52) $$

Secondly, following Watamura$^{[12]}$ we assume that $A^{I}$
have similar properties like $\eta^{I}$. Hereafter, we neglect
the wedge sign $\wedge$ for simplicity.
$$\begin{array}{l}
\Delta_{L}(A^{I})~=~M_{~J}^{I}~\otimes A^{J} \\
({\cal P}_{Si})^{IJ}_{~~KL}~\left(A^{K}~A^{L}\right)~=~0,~~~~
\left(\hat{{\cal P}}_{Ad}\right)^{ij}_{~~k\ell}~\left(A^{k}~
A^{\ell} \right)~=~A^{i} ~A^{j} \\
\displaystyle {ig \over \lambda}~\left(A^{0}~A^{i}
{}~+ ~A^{i} ~A^{0} \right)
{}~=~(\lambda^{2}+2)^{-1}~{\bf C}_{jk}^{~~i}~A^{j}~A^{k}
\end{array} \eqno (2.53) $$

{}From the consistent conditions, $d\eta^{J}$ and $dA^{J}$ have to
satisfy another relation:
$$({\cal P}_{Ad})^{IJ}_{~~KL}~\left(d\eta^{K}~d \eta^{L}\right)~=~0,~~~
({\cal P}_{Ad})^{IJ}_{~~KL}~\left(dA^{K}~dA^{L}\right)~=~0 \eqno (2.54)$$

\noindent
The explicit relations are given in Appendix.

Thirdly, the gauge potential is introduced in the covariant derivative.
The covariant condition of the covariant derivative in the BRST
transformation requires:
$$\delta A^{I}~=~d \eta^{I}~+~\displaystyle { ig \over \lambda}~
\left( \eta^{0}~A^{I}~+~A^{I}~\eta^{0} \right)
 \eqno (2.55) $$

Fourthly, discuss two different fields $X^{J}$ and $Y^{K}$ with
indices $n_{x}$ and $n_{y}$, respectively. For definiteness
assume  $n_{x}>n_{y}$. The consistent condition requires the
following commutative relations:
$$(-1)^{n_{x}n_{y}}~X^{I}~Y^{J}~=~Y^{K}~\left(X^{I}~*~L^{J}_{K}\right)
{}~=~\Lambda^{IJ}_{~~KL}~Y^{K}~X^{L}  \eqno (2.56)$$

\noindent
{}From (2.31) we have
$$\begin{array}{c}
(-1)^{n_{x}n_{y}}~X^{0}~Y^{J}~=~Y^{J}~X^{0} \\
\{ig/\lambda\}~\left( Y^{0}X^{I}~-~(-1)^{n_{x}n_{y}}~X^{I}Y^{0}\right)
{}~=~Y^{J}X^{K}~{\bf C}_{JK}^{~~I}  \end{array} \eqno (2.57) $$

At last, the gauge fields $F^{J}$ satisfy:
$$\begin{array}{c}
F^{J}~=~dA^{J}~+~ig \lambda^{-1}~\left(
A^{0}~A^{J}~+~A^{J}~A^{0}\right)\\
F^{0}~=~dA^{0},~~~~F^{i}~=~dA^{i}~+~(\lambda^{2}+2)^{-1}
{}~{\bf C}_{jk}^{~~i}~A^{j}~A^{k} \\
F^{I}~\eta^{J}~=~\eta^{K}~F^{L}~\Lambda^{IJ}_{~~KL}
 \end{array} \eqno (2.58) $$
$$\begin{array}{rl}
\delta F^{I}&=~ig \lambda^{-1}
{}~\left( \eta^{0}~F^{I}~-~F^{I}~\eta^{0}\right)
{}~=~\eta^{J}~F^{K}~{\bf C}_{JK}^{~~I}\\
\delta F^{0}&=~0,~~~~
\delta F^{i}~=~-~ig\lambda~\eta^{0}~F^{i}~+~{\bf C}_{jk}^{~~i}
{}~\eta^{j}~F^{k} \end{array} \eqno (2.59) $$
$$\begin{array}{rl}
d F^{I}&=~-~ig \lambda^{-1}
{}~\left( A^{0}~F^{I}~-~F^{I}~A^{0}\right) \\
&=~-~ig \lambda^{-1}
{}~\left( A^{0}~dA^{I}~-~dA^{I}~A^{0}\right) \\
&=~-~A^{J}~dA^{K}~{\bf C}_{JK}^{~~I} \\
d F^{0}&=~0 ,~~~~
d F^{i}~=~ig\lambda~A^{0}~dA^{i}~-~{\bf C}_{jk}^{~~i}
{}~A^{j}~dA^{k} \end{array} \eqno (2.60) $$

\vspace{20mm}
\noindent
{\bf 3. $q$-Deformed Chern Class}

It is easy to understand that the second $q$-deformed Chern class has the
following form$^{[11]}$:
$$P~=~Tr_{q}~(F~,~F)~\equiv~F^{I}~F^{J}~g_{IJ} \eqno (3.1) $$

\noindent
where we omit the possible constant factor in $P$. The
$q$-deformed Killing form $g_{IJ}$ is chosen from the condition:
$$\delta P~=~0,~~~~d P~=~0  \eqno (3.2) $$

\noindent
In fact, from (2.58) and (2.59) we have:
$$\begin{array}{rl}
\delta P&=~\left\{\delta F^{R}~F_{S}~+~F^{R}~ \delta F^{S}\right\}
{}~g_{RS} \\
&=~ig \lambda^{-1}~\left\{\eta^{0}~F^{R}~F^{S}~-~
F^{R}~F^{S}~\eta^{0}\right\}~g_{RS}\\
&=~ig \lambda^{-1}~\eta^{I}~F^{J}~F^{K}~
\left\{\delta^{0}_{I}~g_{JK}~-~
\Lambda^{S0}_{~~TK}~\Lambda^{RT}_{~~IJ}~g_{RS}\right\} \end{array} $$

\noindent
namely, $g_{IJ}$ has to satisfy:
$$\delta^{0}_{I}~g_{JK}~=~
\Lambda^{S0}_{~~TK}~\Lambda^{RT}_{~~IJ}~g_{RS} \eqno (3.3) $$

\noindent
It is easy to check that the $g_{IJ}$ satisfying (3.3) also guarantees
$dP=0$. From (A.5) a product of four adjoint components of $A^{j}$
must be vanishing, so that:
$$\begin{array}{rl}
P&=~Tr_{q}(dA^{I}~,~dA^{J})\\
&~~~~+~ig \lambda^{-1}~
\left(Tr_{q}\left\{dA~,~(A^{0}A+AA^{0})\right\}~+~
Tr_{q}\left\{(A^{0}A+AA^{0})~,~dA \right\} \right)\\
dP&=~ig \lambda^{-1}~\left(
Tr_{q}\left\{dA~,~(dA^{0}A-A^{0}dA+dAA^{0}-AdA^{0})\right\} \right.\\
&\left.~~~~+~
Tr_{q}\left\{(dA^{0}A-A^{0}dA+dAA^{0}-AdA^{0})~,~dA \right\} \right)\\
&=~-~ig \lambda^{-1}~\left(
A^{0}~dA^{R}~dA^{S}-dA^{R}~dA^{S}~A^{0} \right)~g_{RS}\\
&=~-~ig \lambda^{-1}~A^{I}~dA^{J}~dA^{K}~
\left\{\delta^{0}_{I}~g_{JK}~-~
\Lambda^{S0}_{~~TK}~\Lambda^{RT}_{~~IJ}~g_{RS}\right\} \end{array} $$

{}From (2.31) we can solve (3.3) as follows.

a) When $J=0$, $I=i\neq 0$ and $K=k\neq 0$, we have:
$$g_{0\ell}~=~0 \eqno (3.4a) $$

b) When $K=0$, $I=i\neq 0$ and $J=j\neq 0$, we have:
$$g_{\ell 0}~=~0 \eqno (3.4b) $$

c) When $I=\pm$ and $J=K=3$, we have:
$$g_{\pm 3}~+~g_{3\pm}~=~0 $$

\noindent
However, when $J=\pm$ and $I=K=3$, we have:
$$\mp~q^{\mp 3}~g_{\pm 3}~+~\lambda ~g_{3\pm}~=~0 $$

\noindent
Thus, we obtain:
$$g_{\pm 3}~=~g_{3\pm}~=~0 \eqno (3.4c) $$

d) When $I=J=\pm$ and $K=3$, we have:
$$g_{++}~=~g_{--}~=~0  \eqno (3.4d) $$

e) When $I=- J=\pm$ and $K=3$, we have:
$$g_{+-}~=~q~g_{33},~~~~g_{-+}~=~q^{-1}~g_{33} \eqno (3.4e) $$

\noindent
For the rest cases there is no new restriction on $g_{IJ}$. It is
worthy to notice that the singlet and adjoint components of $g_{IJ}$
are separated from each other, although those components are
mixed in the commutative relations in ${\cal B}$.

We are going to choose the $q$-deformed Killing form $g_{IJ}$ such that
$g_{IJ}$ satisfies (3.4) and is the analogue of the classical Killing
form. Usually, the quantum trace is related to double $\kappa$
operator$^{[8]}$. Therefore, we choose the $q$-deformed Killing form
as follows:
$$g_{IJ}~=~D^{R}_{~S}~{\bf C}_{IT}^{~~S}~{\bf C}_{JR}^{~~T}
{}~=~D^{r}_{~s}~{\bf C}_{It}^{~~s}~{\bf C}_{Jr}^{~~t}
\eqno (3.5) $$

\noindent
where the summed indices cannot equal to zero owing to (2.45).
{}From (2.11) and (2.45) the non-vanishing components of $g_{IJ}$ are:
$$g_{00}~=~-g^{2}~\lambda^{2}~[3],~~~~q^{-1}~g_{+-}~=~q~g_{-+}~=~g_{33}
{}~=~-g^{2}~(\lambda^{2}+2) \eqno (3.6) $$

\noindent
Obviously, both $g_{IJ}$ and $g_{ij}$ (without $0$ subscript)
satisfy (3.4). From an equivalent condition, Castellani$^{[11]}$
introduced another $q$-deformed Killing form
with an additional parameter $r$.

{}From the $q$-deformed Killing form we may define two kinds
of the second $q$-deformed Chern Class:
$$\begin{array}{rl}
P&=~\langle~F~,~F~\rangle~\equiv~F^{i}~F^{j}~g_{ij} \\
&=~-~g^{2}(\lambda^{2}+2)~\left(F^{3}~F^{3}~+~q~F^{+}~F^{-}~+~
q^{-1}F^{-}~F^{+}\right) \\
\hat{P}&=~Tr_{q}~(F~,~F)~-~P~\equiv~F^{I}~F^{J}~g_{IJ}~-~P \\
&=~-g^{2}~\lambda^{2}~[3]~F^{0}~F^{0}
 \end{array} \eqno (3.7) $$

\noindent
where $P$ and $\hat{P}$ contain only the components of the adjoint
representation and the singlet of $F^{j}$, respectively.

More generally, we define:
$$g_{IJK}~=~D^{t_{1}}_{~t_{2}}~{\bf C}_{It_{3}}^{~~t_{2}}~
{\bf C}_{Jt_{4}}^{~~t_{3}}~{\bf C}_{Kt_{1}}^{~~t_{4}},~~~~
g_{IJKL}~=~D^{t_{1}}_{~t_{2}}~{\bf C}_{It_{3}}^{~~t_{2}}~
{\bf C}_{Jt_{4}}^{~~t_{3}}~{\bf C}_{Kt_{5}}^{~~t_{4}}
{}~{\bf C}_{Lt_{1}}^{~~t_{5}}
\eqno (3.8) $$

\noindent
The non-vanishing components of $g_{IJK}$ are:
$$\begin{array}{l}
g_{000}~=~ig^{3}\lambda^{3}[3],~~~~
g_{0jk}~=~g_{j0k}~=~g_{jk0}~=~-ig\lambda g_{jk}\\
\lambda^{-1}g_{333}~=~-~g_{3+-}~=~g_{3-+}~=~q^{-2}~g_{+3-}
{}~=~-~q^{2}~g_{-3+}\\
{}~~~~~~~~=~-~g_{+-3}~=~g_{-+3}~=~ig^{3}(\lambda^{2}+1)
\end{array} \eqno (3.9) $$

Generalizing (3.1) we define:
$$\begin{array}{l}
Tr_{q}(X_{1}~,~X_{2},~\cdots~,~X_{m})~=~X_{1}^{J_{1}}~X_{2}^{J_{2}}
\cdots X_{m}^{J_{m}}~g_{J_{1}J_{2}\cdots J_{m}} \\
\langle X_{1}~,~X_{2},~\cdots~,~X_{m}\rangle~=~X_{1}^{j_{1}}~X_{2}^{j_{2}}
\cdots X_{m}^{j_{m}}~g_{j_{1}j_{2}\cdots j_{m}}
 \end{array} \eqno (3.10) $$

\noindent
where
$$\begin{array}{rl}
\langle X~,~Y \rangle&=~-~g^{2}(\lambda^{2}+2)~\left\{
X^{3}~Y^{3}~+~q~X^{+}~Y^{-}~+~q^{-1}~X^{-}~Y^{+}\right\}\\
\langle X~,~Y~,~Z \rangle&=~ig^{3}(\lambda^{2}+1)~\left\{
\lambda X^{3}Y^{3}Z^{3}~-~X^{3}Y^{+}Z^{-}~+~
X^{3}Y^{-}Z^{+} \right.\\
&~~~\left.~+~q^{2}X^{+}Y^{3}Z^{-}~-~q^{-2}X^{-}Y^{3}Z^{+}~-~
X^{+}Y^{-}Z^{3}~+~X^{-}Y^{+}Z^{3} \right\}\\
Tr_{q}(X~,~Y)&=~-~g^{2}\lambda^{2}[3]~X^{0}~Y^{0}~+~\langle X~,~Y \rangle \\
Tr_{q}(X~,~Y~,~Z)&=~ig^{3}\lambda^{3}[3]~X^{0}~Y^{0}~Z^{0}~-~ig\lambda
{}~\left\{X^{0}~\langle Y~,~Z \rangle \right. \\
&\left.~~~~+~\langle X~Y^{0}~,~Z \rangle ~+~\langle X~,~Y \rangle
{}~Z^{0}\right\}~+~\langle X~,~ Y~,~Z \rangle
 \end{array} \eqno (3.11) $$

\noindent
where $X^{J}$, $Y^{J}$ and $Z^{J}$ are fields $\eta^{J}$,
$d\eta^{J}$, $A^{J}$ or $dA^{J}$ in the BRST algebra ${\cal B}$.
Their indices are denoted by $n_{x}$,
$n_{y}$ and $n_{z}$, respectively. In (3.10) and (3.11) the fields
can also be replaced by, for example, $XY^{0}$, $Y^{0}X$ or $F$.
It is easy to show the following relations from the explicit values
of the $q$-deformed Killing forms:
$$\begin{array}{l}
\Lambda^{RS}_{~~IJ}~g_{RS}~=~g_{IJ},~~~~
\Lambda^{rs}_{~~ij}~g_{rs}~=~g_{ij} \\
f_{rs}^{i}~g_{rs}~=~0,~~~~
\left(\hat{{\cal P}}_{Ad}\right)^{rs}_{~~ij}~g_{rs}~=~0 \\
\left(\hat{{\cal P}}_{Ad}\right)^{rs}_{~~ij}~g_{rsk}~=~g_{ijk},~~~~
\left(\hat{{\cal P}}_{Ad}\right)^{rs}_{~~jk}~g_{irs}~=~g_{ijk}
\end{array} \eqno (3.12) $$

Now, we are going to derive four important recursive
relations. Assume $X^{J}$ and $Y^{J}$ are two different fields
in ${\cal B}$ with the indices $n_{x}>n_{y}$. $Z^{J}$ denotes a
field in ${\cal B}$, or a field multiplying by the zero component
of this or another field. Since the following relations are
linear for the field $Z^{J}$, $Z^{J}$ can also be replaced by their
linear combination, for example, $F^{J}$.

{}From (2.56) and (2.31) we have:
$$\begin{array}{rl}
(-1)^{n_{x}n_{y}}~X^{i}~Y^{0}&=~Y^{J}~X^{K}~\Lambda^{i0}_{~~JK}\\
&=~(\lambda^{2}+1)~Y^{0}X^{i}~-~(ig)^{-1}\lambda~Y^{j}~X^{k}~
{\bf C}_{jk}^{~~i} \end{array} $$

\noindent
Due to (2.50),
$$(-1)^{n_{x}n_{y}}~X^{i}Y^{0}~{\bf C}_{is}^{~~r}
{}~=~(\lambda^{2}+1)~Y^{0}X^{i}~{\bf C}_{is}^{~~r}
{}~-~\displaystyle {\lambda(\lambda^{2}+2) \over ig(\lambda^{2}+1)}~
Y^{u}X^{v}~\left(\hat{{\cal P}}_{Ad}\right)^{jk}_{~~uv}~
{\bf C}_{jt}^{~~r}{\bf C}_{ks}^{~~t}  $$

\noindent
Now, according to the definition (3.10) and (3.12), we have:
$$\begin{array}{l}
(-1)^{n_{x}n_{y}}~\langle Z~,~X \rangle Y^{0}
{}~=(\lambda^{2}+1)~\langle Z~,~Y^{0}X \rangle
{}~-~\displaystyle {\lambda(\lambda^{2}+2) \over
ig(\lambda^{2}+1)}~\langle Z~,~Y~,~X \rangle \\
(-1)^{n_{x}n_{y}}~\langle XY^{0}~,~Z \rangle
{}~=(\lambda^{2}+1)~Y^{0} \langle X~,~Z \rangle
{}~-~\displaystyle {\lambda(\lambda^{2}+2) \over
ig(\lambda^{2}+1)}~\langle Y~,~X~,~Z \rangle \\
\end{array} \eqno (3.13) $$

\noindent
Similarly, from (2.50) and (2.53) we have:
$$\begin{array}{rl}
\{ig/\lambda\}~\left( A^{0}A^{i}~+~A^{i}A^{0}\right)~{\bf C}_{is}^{~~r}
&=~(\lambda^{2}+2)^{-1}~A^{j}A^{k}~
{\bf C}_{jk}^{~~i}{\bf C}_{is}^{~~r}\\
&=~(\lambda^{2}+1)^{-1}~
A^{j}A^{k}~{\bf C}_{jt}^{~~r}{\bf C}_{ks}^{~~t} \end{array}  $$

\noindent
The equation holds for $\eta^{J}$, too. Thus, we have:
$$\begin{array}{l}
\{ig/\lambda\}~\langle \cdots~,~Z_{1}~,~(A^{0}A+AA^{0})~,~Z_{2}
{}~,~\cdots \rangle \\
{}~~~~~~=~(\lambda^{2}+1)^{-1}~\langle \cdots~,~Z_{1}~,
{}~A~,~A~,~Z_{2}~,~\cdots \rangle \\
\{ig/\lambda\}~\langle \cdots~,~Z_{1}~,~(\eta^{0}\eta+\eta \eta^{0})~,~Z_{2}
{}~,~\cdots \rangle \\
{}~~~~~~=~(\lambda^{2}+1)^{-1}~\langle \cdots~,~Z_{1}~,
{}~\eta~,~\eta~,~Z_{2}~,~\cdots \rangle
\end{array} \eqno (3.14) $$

{}From (2.49) and (2.57) we obtain:
$$\begin{array}{l}
(-1)^{n_{x}n_{y}}~X^{I}Y^{J}~{\bf C}_{It}^{~~r}{\bf C}_{Js}^{~~t}
{}~=~Y^{K}X^{L}~\Lambda^{IJ}_{~~KL}~{\bf C}_{It}^{~~r}{\bf C}_{Js}^{~~t}\\
{}~~~~=~Y^{K}X^{L}~\left\{ {\bf C}_{Kt}^{~~r}{\bf C}_{Ls}^{~~t}
{}~-~{\bf C}_{KL}^{~~t}{\bf C}_{ts}^{~~r} \right\} \\
{}~~~~=~Y^{K}X^{L}~{\bf C}_{Kt}^{~~r}{\bf C}_{Ls}^{~~t}
{}~-~\{ig/\lambda\}~\left(Y^{0}X^{j}~-~(-1)^{n_{x}n_{y}}~X^{j}Y^{0}\right)~
{\bf C}_{js}^{~~r}  \end{array} $$

\noindent
Then,
$$\begin{array}{l}
\left\{(-1)^{n_{x}n_{y}}~X^{i}Y^{j}~-~Y^{i}X^{j}\right\}
{}~{\bf C}_{it}^{~~r}{\bf C}_{js}^{~~t}\\
{}~~~~=~\left\{(-1)^{n_{x}n_{y}}~X^{I}Y^{J}~-~Y^{I}X^{J}\right\}
{}~{\bf C}_{It}^{~~r}{\bf C}_{Js}^{~~t} \\
{}~~~~~~~~+~ig\lambda~\left\{(-1)^{n_{x}n_{y}}~X^{j}Y^{0}-Y^{0}X^{j}\right\}
{}~{\bf C}_{js}^{~~r}\\
{}~~~~=~-~ig\lambda^{-1}(\lambda^{2}+1)~
\left\{Y^{0}X^{j}~-~(-1)^{n_{x}n_{y}}~X^{j}Y^{0}\right\}~
{\bf C}_{js}^{~~r} \end{array} $$

\noindent
So we have,
$$\begin{array}{l}
(-1)^{n_{x}n_{y}}~\langle \cdots~,~Z_{1}~,~X~,~Y~,~Z_{2}
{}~,~\cdots \rangle\\
{}~~~~=~\langle \cdots~,~Z_{1}~,~Y~,~X~,~Z_{2}~,~\cdots \rangle \\
{}~~~~~~~~-~ig\lambda^{-1}(\lambda^{2}+1)~
\langle \cdots~,~Z_{1}~,~(Y^{0}X-(-1)^{n_{x}n_{y}}XY^{0})~,~Z_{2}~,~
\cdots \rangle \end{array} \eqno (3.15) $$

\noindent
By making use of (3.13) we have:
$$\begin{array}{l}
(-1)^{n_{x}n_{y}}~\langle Z~,~X~,~Y \rangle
{}~=~ig\lambda(\lambda^{2}+1)~\langle Z~,~
Y^{0}X \rangle ~-~(\lambda^{2}+1)~
\langle Z~,~Y~,~X \rangle \\
(-1)^{n_{x}n_{y}}~\langle X~,~Y~,~Z \rangle
{}~=~ig\lambda(\lambda^{2}+1)~\langle Y^{0}X~,~Z \rangle
{}~-~(\lambda^{2}+1)~\langle Y~,~X~,~Z \rangle
\end{array} \eqno (3.16) $$

{}From these four recursive relations and (3.12), we can derive
following useful formulas. First of all, from (3.12) we have:
$$\langle \eta~,~ \eta \rangle~=~\langle A~,~ A \rangle~=~0 ,~~~~
(-1)^{n_{x}n_{y}}~\langle X~,~ Y \rangle~=~\langle Y~,~X \rangle
 \eqno (3.17) $$

{}From (3.14) we have:
$$\begin{array}{rl}
\langle F~,~ A \rangle&=~\langle A~,~ F \rangle~=~\langle A~,~ dA \rangle
{}~+~\displaystyle {ig \over \lambda}~\langle A~,~(A^{0}A+AA^{0}) \rangle\\
&=~\langle A~,~ dA \rangle ~+~(\lambda^{2}+1)^{-1}
{}~\langle A~,~A~,~A \rangle
\end{array}\eqno (3.18) $$

If the fields $Y^{J}$ and $Z^{J}$ are not the field $\eta^{J}$, we have:
$$\langle Y~,~Z \rangle~\eta^{0}~=~(-1)^{n_{y}+n_{z}}~
\eta^{0}~ \langle Y~,~Z \rangle \eqno (3.19) $$

Let $Z^{J}$ be a field in ${\cal B}$ with the index $n_{z}$, and
let $X^{J}$ denote the field $\eta^{J}$ or $A^{J}$ with the index
$n_{x}=-1$ or $1$, respectively.  From the recursive relations
we obtain that if $n_{z}<n_{x}$:
$$\begin{array}{l}
\langle X~,~X~,~Z \rangle ~=~\langle Z~,~X~,~X \rangle \\
(-1)^{n_{x}n_{z}}~\langle X~,~Z~,~X \rangle
{}~=~-~(\lambda^{2}+1)~\langle Z~,~X~,~X \rangle \\
\langle X~,~X~,~Z \rangle~-~(-1)^{n_{x}n_{z}} \langle X~,~Z~,~X \rangle
{}~+~\langle Z~,~X~,~X \rangle \\
{}~~~~=~[3]~\langle Z~,~X~,~X \rangle
\end{array} \eqno (3.20) $$

\noindent
If $n_{z}>n_{x}$, we have:
$$\begin{array}{l}
\langle Z~,~X~,~X \rangle~=~ig\lambda(\lambda^{2}+1)(\lambda^{2}+2)~X^{0}
\langle X~,~Z \rangle ~+(\lambda^{2}+1)~\langle X~,~X~,~Z \rangle\\
(-1)^{n_{x}n_{z}}~\langle X~,~Z~,~X \rangle~=~-~ig\lambda(\lambda^{2}+1)~X^{0}
\langle X~,~Z \rangle ~-~\langle X~,~X~,~Z \rangle\\
{}~~~~~~~~=~-~(\lambda^{2}+2)^{-1}~\left\{ \langle X~,~X~,~Z \rangle
{}~+~\langle Z~,~X~,~X \rangle  \right\} \\
\langle X~,~X~,~Z \rangle~-~(-1)^{n_{x}n_{z}} \langle X~,~Z~,~X \rangle
{}~+~\langle Z~,~X~,~X \rangle \\
{}~~~~~~~~=~-~(-1)^{n_{x}n_{z}}~[3]~\langle X~,~Z~,~X \rangle
\end{array} \eqno (3.21) $$

At last, through direct calculation we have:
$$\langle A~,~A~,~A \rangle~\eta^{0}~=~
-~\eta^{0}~ \langle A~,~A~,~A \rangle
{}~=~ig^{3}(\lambda^{2}+1)(\lambda^{2}+2)~\eta^{0}A^{3}A^{+}A^{-}
 \eqno (3.22) $$

\vspace{20mm}
\noindent
{\bf 4. $q$-Deformed Chern-Simons}

In the classical case Zumino $^{[13]}$ introduced a homotopy operator
$k$ to compute the Chern-Simons. Now we generalize his method to
compute the $q$-deformed Chern-Simons.

Introduce a $q$-deformed homotopy operator $k$ that is nilpotent and
satisfies the Leibniz rule in the graded sense for the index $n$:
$$k^{2}~=~0,~~~~d~k~+~k~d~=~{\bf 1} \eqno (4.1) $$

\noindent
If $k$ exists, we can compute the $q$-deformed Chern-Simons $Q(A)$
from $q$-deformed Chern class:
$$\begin{array}{c}
P~=~(d~k~+~k~d)~P~=~d\left(k~P \right)~=~d~Q(A) \\
Q(A)~=~k~P \end{array} \eqno (4.2) $$

\noindent
where we used (3.2).

Introduce a real parameter $t$, $0\leq t \leq 1$. When $t$ changes
from $0$ to $1$, the gauge potentials $A^{J}_{t}$ change from $0$ to
$A^{J}$:
$$\begin{array}{rl}
A^{J}_{t}&=~t~A^{J}\\
F_{t}^{J}&=~t~dA^{J}~+~\displaystyle {ig~t^{2}
\over \lambda}\left(A^{0}~A^{J}~+~A^{J}~A^{0}\right)\\
&=~t~F^{J}~+~\displaystyle {ig~(t^{2}-t)
\over \lambda}\left(A^{0}~A^{J}~+~A^{J}~A^{0}\right)
\end{array} \eqno (4.3) $$

Due to our definition (2.3) for $q$-number we have to a little change
the usual definition$^{[18]}$ for the $q$-deformed derivative and
the $q$-deformed integral. Define the $q$-deformed derivative along $t$ by:
$$\displaystyle {\partial \over \partial_{q} t}~f(t)
{}~=~\displaystyle {f(qt)~-f(q^{-1}t) \over t~(q~-~q^{-1})} \eqno (4.4)$$

\noindent
satisfying the $q$-deformed Leibniz rule:
$$\displaystyle {\partial \over \partial_{q} t}~f(t)g(t)~=~
\displaystyle {\partial f(t)\over \partial_{q} t}~g(qt)~+~
f(q^{-1}t)~\displaystyle {\partial g(t) \over \partial_{q} t}
\eqno (4.5) $$

The $q$-deformed integral is defined by:
$$\displaystyle \int_{0}^{t_{0}}~d_{q}t~f(t)~=~t_{0}(1-q^{2})~
\displaystyle \sum_{k=0}^{\infty}~q^{2k}~f(q^{2k+1}t_{0}) \eqno (4.6) $$

\noindent
At least for a polynomial, the $q$-deformed integral is the inverse
of $q$-deformed derivative. For example,
$$\displaystyle {\partial \over \partial_{q} t}~t^{m}~=~[m]~t^{m-1},~~~~
\displaystyle \int_{0}^{t_{0}}~d_{q}t~t^{m-1}~=~t_{0}^{m}/[m]$$

Now, define the $q$-deformed Lie derivative $\hat{\delta}_{q}$
along $t$ in the gauge space:
$$\hat{\delta}_{q}~\equiv~d_{q}t~\displaystyle {\partial
\over \partial_{q} t} \eqno (4.7) $$

\noindent
and the $q$-deformed operator $\ell_{t}$ that satisfies the
$q$-deformed Leibniz rule in the graded sense for the index $n$:
$$\begin{array}{l}
\ell_{t} A_{t}^{J}~=~0,~~~~
\ell_{t} F_{t}^{J}~=~\hat{\delta}_{q} A^{J}_{t}~=~d_{q}t~A^{J} \\
\ell_{t}\left\{f(t)g(t)\right\}
{}~=~\left\{\ell_{t} f(t)\right\}~g(qt)~+~(-1)^{n}~f(q^{-1}t)
{}~\left\{\ell_{t} g(t)\right\}
\end{array} \eqno (4.8) $$

\noindent
where $f$ has index $n$.

It is easy to check that for all formal polynomials (vanishing at
$F^{J}_{t}=0$ and $A^{J}_{t}=0$) we have
$$\begin{array}{c}
\ell_{t}~\ell_{t}~=~0\\
\ell_{t}~d~+~d~\ell_{t}~=~\hat{\delta}_{q}~=~d_{q}t~\displaystyle {\partial
\over \partial_{q} t} \\
\hat{\delta}_{q}~d~=~d~\hat{\delta}_{q},~~~~
\hat{\delta}_{q}~\ell_{t}~=~\ell_{t}~\hat{\delta}_{q}~
\end{array} \eqno (4.9) $$

\noindent
Comparing it with (4.1) we obtain:
$$k~=~\displaystyle \int_{0}^{1}~\ell_{t} \eqno (4.10) $$

Now, we are able to compute the $q$-deformed Chern-Simons by (4.2):
$$\begin{array}{rl}
\ell_{t} P_{t}&=~\langle \ell_{t}F_{t}~,~F_{qt}\rangle
{}~+~\langle F_{q^{-1}t}~,~\ell_{t}F_{t}\rangle \\
&=~d_{q}t~\left\{ \langle A~,~F_{qt}\rangle~+~\langle
F_{q^{-1}t}~,~A \rangle \right\}\\
&=~d_{q}t~\left\{t[2] \langle A~,~dA \rangle~+~ig\lambda^{-1}t^{2}
(\lambda^{2}+2)~\langle A~,~(A^{0}A+AA^{0}) \rangle \right\}\\
&=~d_{q}t~\left\{ t~[2]~\langle A~,~dA \rangle~+~\displaystyle
{t^{2}(\lambda^{2}+2) \over \lambda^{2}+1}~\langle A~,~A~,~A \rangle
\right\} \end{array} \eqno (4.11) $$

\noindent
where we have used (3.14).
$$\begin{array}{rl}
Q(A)&=~k~P_{t} \\
&=~\langle A~,~dA \rangle~+~ \{ [4]/[6]\}
{}~\langle A~,~A~,~A\rangle \\
&=~\langle A~,~F\rangle~-~\{[2]/[6]\}
{}~\langle A~,~A~,~A\rangle \end{array} \eqno (4.12) $$

\noindent
Similarly, we can compute $\hat{Q}(A)$ from $\hat{P}=d\hat{Q}(A)$:
$$\begin{array}{rl}
\ell_{t} \hat{P}_{t}&=~-~g^{2}\lambda^{2}[3]~\left\{
\left( \ell_{t}F^{0}_{t}\right)~F^{0}_{qt}
{}~+~F^{0}_{q^{-1}t}~\left(\ell_{t}F_{t}\right) \right\} \\
&=~-~g^{2}\lambda^{2}[3]~d_{q}t~\left\{ A^{0}~F^{0}_{qt}~+~
F^{0}_{q^{-1}t}~A^{0} \right\}\\
&=~-~g^{2}\lambda^{2}[3]~d_{q}t~\left\{ t~[2]~A^{0}~dA^{0} \right\}
\end{array} \eqno (4.13) $$
$$\hat{Q}(A)~=~k~\hat{P}_{t}
{}~=~-~g^{2}\lambda^{2}[3]~A^{0}~dA^{0} \eqno (4.14) $$

\noindent
It is easy to prove by direct calculation that:
$$dQ(A)~=~P,~~~~d\hat{Q}(A)~=~\hat{P} \eqno (4.15) $$

\noindent
In fact, the second equation is obvious, and the first one can be
proved in terms of the formulas given in the last section:
$$\begin{array}{rl}
d Q(A)&=~\langle dA~,~dA \rangle  \\
&~~~~+~\{[4]/[6]\}~\left\{ \langle dA~,~A~,~A \rangle ~-~
\langle A~,~dA~,~A \rangle ~+~\langle A~,~A~,~dA \rangle \right\} \\
&=~\langle dA~,~dA \rangle ~+~
(\lambda^{2}+1)^{-1}~\left\{ \langle dA~,~A~,~A \rangle
{}~+~\langle A~,~A~,~dA \rangle \right\} \\
&=~~\langle dA~,~dA \rangle ~+~
\displaystyle {ig\over \lambda}
\left\{\langle dA~,~(A^{0}A+AA^{0})\rangle
{}~+~ \langle (A^{0}A+AA^{0})~,~dA \rangle \right\} \\
&=~\langle F~,~F\rangle~=~P \end{array} $$

We find that the components of the singlet and the adjoint representation
are separated in the $q$-deformed Chern class and in the $q$-deformed
Chern-Simons, although they are mixed in the commutative relations of
BRST algebra.

\vspace{20mm}
\noindent
{\bf 5. $q$-Deformed Cocycle Hierarchy}

Just like those in the classical case$^{[19]}$, the gauge fields $F^{J}$ are
invariant under the transformation:
$$A^{J}~\rightarrow~A^{J}~-~\eta^{J},~~~~d~\rightarrow~d~+~\delta
\eqno (5.1) $$

\noindent
In fact,
$$\begin{array}{rl}
F^{J}&\rightarrow~{\cal F}^{J} \\
&=~(d~+~\delta)~(A^{J}~-~\eta^{J}) \\
&~~~~+~\displaystyle {ig \over \lambda}~
\left\{ (A^{0}~-~\eta^{0})(A^{J}~-~\eta^{J})~+~
(A^{J}~-~\eta^{J})(A^{0}~-~\eta^{0}) \right\}\\
&=~F^{J}~+~\left\{\delta A^{J}~-~d\eta^{J}~-~\displaystyle {ig \over
\lambda}~(\eta^{0}A^{J}~+~A^{J}\eta^{0})\right\} \\
&~~~~-~\left\{\delta\eta^{J}~-~\displaystyle {ig \over \lambda}~
\left(\eta^{0}\eta^{J}~+~\eta^{J}\eta^{0}-A^{0}\eta^{J}~-~\eta^{J}A^{0}
\right) \right\} \\
&=~F^{J} \end{array} $$

Now, under the transformation equation (4.15) becomes:
$$\begin{array}{l}
P~=~(d~+~\delta)~Q(A-\eta) \\
Q(A-\eta)~=~ \langle A-\eta~,~F \rangle~-~\{[2]/[6]\}
{}~\langle A-\eta~,~A-\eta~,~A-\eta\rangle
\end{array} \eqno (5.2) $$

\noindent
Expanding (5.2) by the ghost number, we have:
$$\begin{array}{l}
Q(A-\eta)~=~\omega_{3}^{0}~+~\omega_{2}^{1}
{}~+~\omega_{1}^{2} ~+~\omega_{0}^{3} \\
P~=~d \omega_{3}^{0}~+~\left\{
\delta \omega_{3}^{0}~+~d \omega_{2}^{1}\right\}~+~
\left\{\delta \omega_{2}^{1}~+~d \omega_{1}^{2}\right\}\\
{}~~~~~~~~+~\left\{\delta \omega_{1}^{2}~+~d \omega_{0}^{3}\right\}
{}~+~\delta \omega_{0}^{3}
\end{array} \eqno (5.3) $$

\noindent
where the subscripts denote the degrees of form of the quantities,
and the superscripts denote the ghost numbers. In two sides of (5.3)
the quantities with the same degree of form and the same ghost number
should be equal to each other, respectively:
$$\begin{array}{l}
P~=~d \omega_{3}^{0},~~~~~~
\delta \omega_{3}^{0}~+~d \omega_{2}^{1}~=~0,~~~~~~
\delta \omega_{2}^{1}~+~d \omega_{1}^{2}~=~0 \\
\delta \omega_{1}^{2}~+~d \omega_{0}^{3}~=~0,~~
{}~~~~\delta \omega_{0}^{3}~=~0
\end{array} \eqno (5.4) $$

We are going to derive the explicit forms of those $\omega$,
and simplify them by the formulas given in Section 3.
$$\begin{array}{rl}
\omega_{3}^{0}&=~Q(A)~=~\langle A~,~F\rangle
{}~-~\{[2]/[6]\}~\langle A~,~A~,~A \rangle \\
&=~\langle A~,~dA \rangle
{}~+~\{[4]/[6]\}~\langle A~,~A~,~A \rangle
\end{array} \eqno (5.5) $$
$$\begin{array}{rl}
\omega_{2}^{1}&=~-~\langle \eta~,~dA \rangle ~-~\{ig/\lambda \}~
\langle \eta~,~(A^{0}A+AA^{0}) \rangle \\
&~~~~+~\{[2]/[6]\}~\left\{\langle \eta~,~A~,~A \rangle~+~
\langle A~,~\eta~,~A \rangle~+~\langle A~,~A~,~\eta \rangle
\right\} \\
&=~-~\langle \eta~,~dA\rangle
\end{array} \eqno (5.6) $$
$$\begin{array}{rl}
\omega_{1}^{2}&=~-~\{[2]/[6]\}~\left\{
\langle \eta~,~\eta~,~A \rangle~+~
\langle\eta~,~A~,~\eta \rangle~+~\langle A~,~\eta~,~\eta \rangle
\right\} \\
&=~-~(\lambda^{2}+1)^{-1}~\langle \eta~,~A~,~\eta \rangle
\end{array} \eqno (5.7) $$
$$\begin{array}{l}
\omega_{0}^{3}~=~\{[2]/[6]\}~\langle \eta~,~\eta~,~\eta \rangle
\end{array} \eqno (5.8) $$

\noindent
Similarly, we also have:
$$\begin{array}{rl}
\hat{Q}(A-\eta)&=~-~g^{2}\lambda^{2}[3]~\left(A^{0}~-~\eta^{0}\right)~F^{0}\\
&=~\hat{\omega}_{3}^{0}~+~\hat{\omega}_{2}^{1}\\
\hat{\omega}_{3}^{0}&=~\hat{Q}(A)~=~-~g^{2}\lambda^{2}[3]~A^{0}~F^{0}\\
\hat{\omega}_{2}^{1}&=~g^{2}\lambda^{2}[3]~\eta^{0}~F^{0}\\
\hat{P}~=~d\hat{\omega}_{3}^{0},~&~~
\delta \hat{\omega}_{3}^{0}~+~d \hat{\omega}_{2}^{1}~=~0,~~~~
\delta \hat{\omega}_{2}^{1}~=~0
\end{array} \eqno (5.9) $$

Equation (5.9) is obviously. We are going to prove (5.4) again directly
by the formulas in Section 3.
$$\begin{array}{l}
\delta \omega_{3}^{0}~+~d \omega_{2}^{1} \\
{}~~~~=~\langle \delta A~,~dA \rangle
{}~+~\langle A~,~d(\delta A) \rangle \\
{}~~~~~~~~+~\{[4]/[6]\}~
\left\{ \langle \delta A~,~A~,~A \rangle
{}~-~\langle A~,~\delta A~,~A \rangle
{}~+~\langle A~,~A~,~\delta A \rangle  \right\}\\
{}~~~~~~~~-~\langle d\eta~,~dA\rangle \\
{}~~~~=~\{ig / \lambda\}~\left\{\langle (\eta^{0} A+A\eta^{0})
{}~,~dA\rangle ~+~\langle
A~,~(d\eta^{0} A-\eta^{0} dA+dA\eta^{0}-A d\eta^{0}) \rangle \right\}\\
{}~~~~~~~~+~\{[4]/[6]\}~\left\{\langle d\eta~,~A~,~A \rangle~-~
\langle A~,~d\eta~,~A \rangle~+~\langle A~,~A~,~d\eta \rangle \right\} \\
{}~~~~~~~~+~\displaystyle {ig[4]\over \lambda [6]}~
\left\{~\langle (\eta^{0}A+A\eta^{0})~,~A~,~A \rangle
{}~-~\langle  A~,~(\eta^{0}A+A\eta^{0})~,~A \rangle \right. \\
\left.~~~~~~~~+~\langle  A~,~A~,~(\eta^{0}A+A\eta^{0}) \rangle  \right\} \\
{}~~~~=~\{ig / \lambda\}~ \left\{\eta^{0} \langle A~,~dA\rangle
{}~+~\langle Ad\eta^{0}~,~A \rangle
{}~+~ \langle A~,~dA \rangle~\eta^{0} \right\} \\
{}~~~~~~~~~+~\{[4][3]/[6]\}~\langle d\eta~,~A~,~A \rangle
{}~+~\displaystyle {ig[4]\over \lambda [6]}~
\left\{\eta^{0}\langle A~,~A~,~A \rangle~+~
\langle  A~,~A~,~A \rangle \eta^{0} \right\} \\
{}~~~~=~0 \end{array} $$

$$\begin{array}{l}
\delta \omega_{2}^{1}~+~d \omega_{1}^{2} \\
{}~~~~=~-~\langle \delta \eta~,~dA \rangle
{}~-~\{ig/\lambda\}~\langle \eta~,~(d\eta^{0} A-\eta^{0} dA+dA \eta^{0}
-Ad\eta^{0}) \rangle \\
{}~~~~~~~~-~(\lambda^{2}+1)^{-1}~
\left\{ \langle d \eta~,~A~,~\eta \rangle
{}~-~\langle \eta~,~dA~,~\eta \rangle
{}~+~\langle \eta~,~A~,~d \eta \rangle  \right\}\\
{}~~~~=~-~(\lambda^{2}+1)^{-1}~\langle  \eta~,~\eta~,~dA \rangle
{}~-~\{ig/\lambda\}~\langle \eta d\eta^{0}~,~ A \rangle
{}~+~ig\lambda~\eta^{0} \langle \eta ~,~ dA \rangle \\
{}~~~~~~~~-~\lambda^{2}(\lambda^{2}+1)^{-1}~\langle \eta ~,~\eta~,~ dA \rangle
{}~+~\displaystyle {\lambda^{2}+2 \over \lambda^{2}+1}
{}~\langle \eta ~,~\eta~,~ dA \rangle \\
{}~~~~~~~~+~\{ig/\lambda\}(\lambda^{2}+1)~\langle \eta d\eta^{0}~,~ A \rangle
{}~-~\displaystyle {\lambda^{2}+2 \over \lambda^{2}+1}
{}~\langle \eta~,~ d\eta~,~ A \rangle \\
{}~~~~~~~~-~(\lambda^{2}+1)^{-1}~\left\{
{}~-~\langle \eta~,~ d\eta~,~ A \rangle
{}~+~ig\lambda(\lambda^{2}+1)~\eta^{0}~\langle \eta~,~ dA \rangle \right.\\
\left.~~~~~~~~+~\langle \eta ~,~\eta~,~ dA \rangle
{}~+~ig\lambda(\lambda^{2}+1)~\langle \eta d\eta^{0}~,~ A \rangle
{}~-~(\lambda^{2}+1)~\langle \eta~,~ d\eta~,~ A \rangle \right\}\\
{}~~~~=~0 \end{array} $$

$$\begin{array}{l}
\delta \omega_{1}^{2}~+~d \omega_{0}^{3} \\
{}~~~~=~-~(\lambda^{2}+1)^{-1}~
\left\{ \langle \delta \eta~,~A~,~\eta \rangle
{}~-~\langle \eta~,~\delta A~,~\eta \rangle
{}~+~\langle \eta~,~A~,~\delta \eta \rangle  \right\}\\
{}~~~~~~~~+~\{[2]/[6]\}~
\left\{ \langle d \eta~,~\eta~,~\eta \rangle
{}~-~\langle \eta~,~d \eta~,~\eta \rangle
{}~+~\langle \eta~,~\eta~,~d \eta \rangle  \right\}\\
{}~~~~=~-~(\lambda^{2}+1)^{-1}~\left\{(\lambda^{2}+1)^{-1}
{}~\langle \eta~,~\eta~,~A~,~\eta \rangle
{}~-~\langle \eta~,~d\eta~,~\eta \rangle
\right. \\
\left.~~~~~~~~-~\{ig/\lambda\}~\langle \eta~,~
(\eta^{0}A+A\eta^{0})~,~\eta \rangle
{}~+~(\lambda^{2}+1)^{-1}~
\langle \eta~,~A~,~\eta~,~\eta \rangle \right\} \\
{}~~~~~~~-~\{[2][3]/[6]\}~\langle \eta~,~d\eta~,~\eta \rangle \\
{}~~~~=~0 \end{array} $$

A product of four adjoint components of $\eta^{j}$ must be vanishing,
so we have:
$$\begin{array}{rl}
\delta \omega_{0}^{3}&=~\{ [2]/[6]\}~\left\{
\langle \delta \eta~,~\eta~,~\eta \rangle
{}~-~\langle \eta~,~\delta \eta~,~\eta \rangle
{}~+~\langle \eta~,~\eta~,~\delta \eta \rangle \right\} \\
&=~\{ [2]/[6]\}~\langle \eta~,~ \eta~,~\eta~,~\eta \rangle \\
&=0 \end{array} $$

\vspace{20mm}
\noindent
{\bf 6. $q$-deformed Lagrangian and Yang-Mills equation}

In the present paper the spacetime is the ordinary commutative
Minkowski spacetime. Explicitly writing down the spacetime indices,
we have:
$$A^{J}~=~A^{J}_{\mu}~dx^{\mu},~~~~
F^{J}~=~\displaystyle {1\over 2}~F^{J}_{\mu \nu}~dx^{\mu}
\wedge dx^{\nu} \eqno (6.1) $$

\noindent
It is well known that the metric $g^{\mu \nu}$ in the Minkowski spacetime
can change the covariant index to contravariant index, or vice versa.

Now, the $q$-deformed Lagrangian that is covariant both in the Lorentz
transformation and the $q$-gauge transformation is:
$${\cal L}~=~-~\displaystyle {1\over 4}~\langle F_{\mu \nu}~,~
F^{\mu \nu} \rangle ~=~-~\displaystyle {1\over 4}~(F^{i})_{\mu \nu}~
(F^{j})^{\mu \nu}~g_{ij}  \eqno (6.3) $$

\noindent
We have known that the components of the singlet and the adjoint
representation are separated in the $q$-deformed Chern class, and
obviously in the $q$-deformed Lagrangian. Here we only discuss
the $q$-deformed Lagrangian constructed by the adjoint components.

The $q$-deformed Yang-Mills equation is just the $q$-deformed
Lagrangian equation:
$$\partial_{\mu}~(F^{j})^{\mu \nu}~\left(g_{kj}+g_{jk}\right)
{}~=~(\lambda^{2}+2)^{-1}~\left({\bf C}_{rk}^{~~i}-{\bf C}_{kr}^{~~i}\right)~
\left\{g_{ij}(A^{r})_{\mu}(F^{j})^{\mu \nu}~+~
g_{ji}(F^{j})^{\mu \nu}(A^{r})_{\mu}\right\} \eqno (6.4) $$

\noindent
In component form we have:
$$\begin{array}{rl}
\partial_{\mu}~(F^{3})^{\mu \nu}
&=~\displaystyle {ig[2] \over 2(\lambda^{2}+2)}~\left\{
q^{-1}~(A^{-})_{\mu}~(F^{+})^{\mu \nu}~+~
q~(F^{+})^{\mu \nu}~(A^{-})_{\mu} \right. \\
&\left.~~~~-~
q~(A^{+})_{\mu}~(F^{-})^{\mu \nu}~-~
q^{-1}~(F^{+})^{\mu \nu}~(A^{-})_{\mu} \right\}
\end{array} \eqno (6.5) $$
$$\begin{array}{rl}
\partial_{\mu}~(F^{+})^{\mu \nu}
&=~\displaystyle {ig \over \lambda^{2}+2}~\left\{
\displaystyle {2 \over [2]}~(A^{+})_{\mu}~(F^{3})^{\mu \nu}~+~
\displaystyle {2 \over [2]}~(F^{3})^{\mu \nu}~(A^{+})_{\mu} \right. \\
&\left.~~~~-~q^{-1}~(A^{3})_{\mu}~(F^{+})^{\mu \nu}~-~
q~(F^{+})^{\mu \nu}~(A^{3})_{\mu} \right\}
\end{array} \eqno (6.6) $$
$$\begin{array}{rl}
\partial_{\mu}~(F^{-})^{\mu \nu}
&=~\displaystyle {ig \over \lambda^{2}+2}~\left\{
q~(A^{3})_{\mu}~(F^{-})^{\mu \nu}~+~
q^{-1}~(F^{-})^{\mu \nu}~(A^{3})_{\mu} \right. \\
&\left.~~~~-~\displaystyle {2 \over [2]}~
(A^{-})_{\mu}~(F^{3})^{\mu \nu}~-~\displaystyle {2 \over [2]}~
(F^{3})^{\mu \nu}~(A^{-})_{\mu} \right\}
\end{array} \eqno (6.7) $$

\vspace{2.0cm}
{\bf Acknowledgments}. This work was supported by the National
Natural Science Foundation of China and Grant No. LWTZ-1298 of
Chinese Academy of Sciences.

\vspace{10mm}
\noindent
{\bf Appendix} Some explicit expressions

a) The components of $M^{I}_{~J}$ (see (2.11)):
$$\begin{array}{l}
M^{0}_{~0}~=~{\bf 1},~~~~M^{i}_{~0}~=~M^{0}_{~i}~=~0\\
M^{3}_{~3}~=~{\bf 1}+[2]\beta \gamma,~~~~M^{+}_{~+}~=~\delta \delta,~~~~
M^{-}_{~-}~=~\alpha \alpha\\
M^{3}_{~+}~=~q^{-3/2}[2]^{1/2}~\delta \beta,~~~~
M^{3}_{~-}~=~-~q^{3/2}[2]^{1/2}~\alpha \gamma \\
M^{+}_{~3}~=~q^{3/2}[2]^{1/2}~\gamma \delta,~~~~
M^{-}_{~3}~=~-~q^{-3/2}[2]^{1/2}~\beta \alpha  \\
M^{+}_{~-}~=~-~q^{4}~\gamma \gamma,~~~~
M^{-}_{~+}~=~-~q^{-4}~\beta \beta
\end{array} \eqno (A.1) $$

b) Non-vanishing components of $L^{J}_{~K}\left(T^{a}_{~b}\right)$
(see (2.22)):
$$\begin{array}{l}
L^{0}_{0}(\alpha)~=~L^{0}_{0}(\delta)~=~[4][2]^{-2},~~~~
L^{3}_{3}(\alpha)~=~L^{3}_{3}(\delta)~=~2[2]^{-1},\\
L^{0}_{3}(\alpha)~=~-~L^{3}_{0}(\delta)~=~q\lambda [2]^{-1},~~~~
L^{3}_{0}(\alpha)~=~-~L^{0}_{3}(\delta)~=~q^{-1}\lambda [2]^{-1},\\
L^{+}_{+}(\alpha)~=~L^{-}_{-}(\alpha)~=~
L^{+}_{+}(\delta)~=~L^{-}_{-}(\delta)~=~1, \\
L^{+}_{0}(\beta)~=~L^{+}_{3}(\beta)~=~
L^{0}_{-}(\beta)~=~-~L^{3}_{-}(\beta)~=~q^{3/2} \lambda [2]^{-1/2} \\
L^{-}_{0}(\gamma)~=~L^{-}_{3}(\gamma)~=~
L^{0}_{+}(\gamma)~=~-~L^{3}_{+}(\gamma)~=~q^{-3/2} \lambda [2]^{-1/2}
\end{array} \eqno (A.2) $$

c) Non-vanishing components of $\chi_{J}\left(T^{a}_{~b}\right)$
(see (2.27)):
$$\begin{array}{l}
\chi_{0}(\alpha)~=~\chi_{0}(\delta)~=~\displaystyle {ig(q^{-1}-q^{2})
\over [2](q+1)},~~~~\chi_{3}(\alpha)~=~-~igq [2]^{-1} ,~~~~
\chi_{3}(\delta)~=~igq^{-1} [2]^{-1} \\
\chi_{-}(\beta)~=~-~igq^{3/2} [2]^{-1/2},~~~~
\chi_{+}(\gamma)~=~-~igq^{-3/2} [2]^{-1/2}
\end{array} \eqno (A.3) $$

d) Non-vanishing components of $({\cal P}_{Si})^{IJ}_{~~KL}$ and
$({\cal P}_{Ad})^{IJ}_{~~KL}$ (see (2.39) and (2.41)):
$$\begin{array}{c}
({\cal P}_{Si})^{ij}_{~~k\ell}~=~\delta^{i}_{k}~\delta^{j}_{\ell}
{}~-~2[2]^{-2}~f_{ij}^{n}~f_{k\ell}^{n},~~~
({\cal P}_{Si})^{00}_{~~00}~=~1 \\
({\cal P}_{Ad})^{ij}_{~~k\ell}~=~2[2]^{-2}~f_{ij}^{n}~f_{k\ell}^{n}\\
({\cal P}_{Si})^{0i}_{~~jk}~=~({\cal P}_{Si})^{i0}_{jk}~=
{}~-~({\cal P}_{Ad})^{0i}_{~~jk}~=~-~({\cal P}_{Ad})^{i0}_{jk}~=~
[2]^{-2}~\lambda~f_{jk}^{i} \\
({\cal P}_{Si})^{jk}_{~~0i}~=~({\cal P}_{Si})^{jk}_{~~i0}~=
{}~-~({\cal P}_{Ad})^{jk}_{~~0i}~=~-~({\cal P}_{Ad})^{jk}_{~~i0}~=
{}~-~[2]^{-2}~\lambda~f_{jk}^{i} \\
({\cal P}_{Si})^{0i}_{~~j0}~=~({\cal P}_{Si})^{i0}_{~~0j}~=~
-~({\cal P}_{Ad})^{0i}_{~~j0}~=~-~({\cal P}_{Ad})^{i0}_{~~0j}~=~
(\lambda^{2}+2)[2]^{-2}~\delta^{i}_{j}
\end{array} \eqno (A.4)$$

e) The explicit commutative relations in BRST algebra (see (2.40),
(2.53) and (2.54)):
$$\begin{array}{l}
A^{0}A^{0}~=~A^{+}A^{+}~=~A^{-}A^{-}~=~0,~~~A^{+}A^{-}~=~-~A^{-}A^{+}\\
A^{\pm}A^{3}~=~-~q^{\pm 2}A^{3}A^{\pm},~~~~
A^{0}A^{\pm}~+~A^{\pm}A^{0}~=~\pm~q^{\pm 1}~\lambda~A^{3}A^{\pm} \\
A^{0}A^{3}~+~A^{3}A^{0}~=~\lambda~A^{+}A^{-}~=~~-~A^{3}A^{3}
\end{array} \eqno (A.5) $$
$$\begin{array}{l}
dA^{0}~dA^{J}~=~dA^{J}~dA^{0},~~~~dA^{\pm}~dA^{3}~-q^{\mp 2}~
dA^{3}~dA^{\pm}~=~\mp~q^{\mp 1}~\lambda~dA^{0}~dA^{\pm}\\
dA^{+}~dA^{-}~-~dA^{-}~dA^{+}~=~\lambda~\left(dA^{3}~dA^{3}~
{}~+~dA^{3}~dA^{0}\right) \end{array} \eqno (A.6) $$

\noindent
(A.5) and (A.6) also hold if $A^{J}$ are replaced by $\eta^{J}$.

For two different fields $X^{J}$ and $Y^{J}$ in ${\cal B}$
with indices $n_{x}$ and $n_{y}$, $n_{x}>n_{y}$, respectively,
we have:
$$\begin{array}{l}
(-1)^{n_{x}n_{y}}~X^{0}~Y^{J}~=~Y^{J}~X^{0},~~~~
(-1)^{n_{x}n_{y}}~X^{\pm}~Y^{\pm}~=~Y^{\pm}~X^{\pm}\\
(-1)^{n_{x}n_{y}}~X^{3}~Y^{0}~=~(\lambda^{2}+1)~Y^{0}~X^{3}~
{}~\lambda^{2}~Y^{3}~X^{3}~+~\lambda~Y^{-}~X^{+}
{}~-~\lambda~Y^{+}~X^{-}\\
(-1)^{n_{x}n_{y}}~X^{\pm}~Y^{0}~=~(\lambda^{2}+1)~Y^{0}~X^{\pm}~
\mp~q^{\mp 1}\lambda~Y^{3}~X^{\pm}~\pm~q^{\pm 1}\lambda~Y^{\pm}~X^{3}\\
(-1)^{n_{x}n_{y}}~X^{3}~Y^{3}~=~(1-\lambda^{2})~Y^{3}~X^{3}~-~
\lambda^{2}~Y^{0}~X^{3}~+~\lambda~Y^{+}~X^{-}~-~\lambda~Y^{-}~X^{+}\\
(-1)^{n_{x}n_{y}}~X^{3}~Y^{\pm}~=~Y^{\pm}~X^{3}~\pm~q^{\mp 1}~\lambda~
\left( Y^{3}~X^{\pm}~+~Y^{0}~X^{\pm}\right) \\
(-1)^{n_{x}n_{y}}~X^{\pm}~Y^{3}~=~Y^{3}~X^{\pm}~\mp~q^{\pm 1}~\lambda~
\left( Y^{0}~X^{\pm}~+~Y^{\pm}~X^{3}\right) \\
(-1)^{n_{x}n_{y}}~X^{\pm}~Y^{\mp}~=~Y^{\mp}~X^{\pm}~\pm~\lambda~
\left( Y^{0}~X^{3}~+~Y^{3}~X^{3}\right)
\end{array} \eqno (A.7) $$

\noindent
Owing to (2.58) $F^{I}$ and $\eta^{J}$ satisfy the similar
commutative relations like (A.7), but there is a tail in those
between $F^{I}$ and $A^{J}$:
$$\begin{array}{l}
F^{0}A^{J}~=~A^{J}F^{0},~~~~
F^{\pm}A^{\pm}~=~A^{\pm}F^{\pm}\\
F^{3}A^{0}~=~(\lambda^{2}+1)~A^{0}F^{3}~
+~\lambda^{2}~A^{3}F^{3}~+~\lambda~A^{-}F^{+}
{}~-~\lambda~A^{+}F^{-}~-~ig\lambda^{2}~A^{0}A^{+}A^{-}\\
F^{\pm}A^{0}~=~(\lambda^{2}+1)~A^{0}F^{\pm}~
\mp~q^{\mp 1}\lambda~A^{3}F^{\pm}~\pm~q^{\pm 1}\lambda~A^{\pm}F^{3}
{}~\mp~igq^{\pm}\lambda^{2}~A^{0}A^{3}A^{\pm}\\
F^{3}A^{3}~=~(1-\lambda^{2})~A^{3}F^{3}~-~
\lambda^{2}~A^{0}F^{3}~+~\lambda~A^{+}F^{-}~-~\lambda~A^{-}F^{+}
{}~+~ig\lambda^{2}~A^{0}A^{+}A^{-}\\
F^{3}A^{\pm}~=~A^{\pm}F^{3}~\pm~q^{\mp 1} \lambda~\left( A^{3}F^{\pm}
{}~+~A^{0}F^{\pm}\right)~-~ig \lambda~A^{0}A^{3}A^{\pm} \\
F^{\pm}A^{3}~=~A^{3}F^{\pm}~\mp~q^{\pm 1} \lambda~\left( A^{0}F^{\pm}
{}~+~A^{\pm}F^{3}\right)~+~igq^{\pm 2}\lambda~A^{0}A^{3}A^{\pm} \\
F^{\pm}A^{\mp}~=~A^{\mp} F^{\pm}~\pm \lambda~\left( A^{0}F^{3}
{}~+~A^{3}F^{3}\right)~\mp~ig\lambda~A^{0}A^{+}A^{-}
\end{array} \eqno (A.8) $$


\newpage
\begin{thebibliography}{99}
\bibitem{1} Yu. I. Manin, {\it Quantum Groups and Non-Commutative
Geometry}, Montreal Univ. preprint, CRM-1561, 1988.
\bibitem{2} S. L. Woronowicz, {\it Commun. Math. Phys}. {\bf 111}(1987)613;
{\bf 122}(1989)125.
\bibitem{3} J. Wess and B. Zumino, {\it Nucl. Phys}. (Proc. Suppl.)
{\bf B18}(1990)302.
\bibitem{4} D. Bernard, Quantum Lie Algebras and Differential Calculus
on Quantum Groups -- Proc. 1990 Yukawa Int. Seminar (kyoto); {\it Phys.
Lett}. {\bf B260}(1991)389.
\bibitem{5} B. Jue\~{c}o, {\it Lett. Math. Phys}. {\bf 22}(1991)177.
\bibitem{6} B. Zumino, {\it Introduction to the differential geometry of
quantum groups}, LBL-31432 and UCB-PTH-62-91, Notes of a plenary talk
given at the 10th IAMP Conf., Leipzig, 1991.
\bibitem{7} U. Carow-Watamura, M. Schlieker, S. Watamura and
W. Weich, {\it Commun. Math. Phys}. {\bf 142}(1991)605.
\bibitem{8} A. P. Isaev and Z. Popowicz, {\it $q$-trace for the quantum
groups and $q$-deformed Yang-Mills theory}, Wroclaw preprint ITP UWr 786/91;
I. Ya. Aref'eva and I. V. Volovich, {\it Phys. Lett}.
{\bf B264}(1991)62; {\it Mod. Phys. Lett}. {\bf A6}(1991)893.
\bibitem{9} X. D. Sun and S. K. Wang, {\it Bicovariant differential calculus
on quantum group $GL_{q}(n)$}, Worldlab-Beijing preprint CCAST-92-04, 1992.
\bibitem{10} P. Aschieri, L. Castellani, {\it Inter. J. Mod. Phys}.
{\bf A8}(1993)1667.
\bibitem{11} L. Castellani, {\it Phys. Lett}. {\bf B292}(1992)93;
{\it $U_{q}(N)$ Gauge Theories}, Preprint, DFTT-74/92, 1992.
\bibitem{12} S. Watamura, {\it Quantum deformation of BRST algebra},
Preprint HD-THEP-92-39, 1992.
\bibitem{13} B. Zumino, {\it Chiral Anomalies and Differential Geometry},
Lectures given at Les Houches, Preprint, LBL-16747, 1983; J. Ma\~{n}es,
R. Stora and B. Zumino, {\it Commun. Math. Phys}. {\bf 102}(1985)157.
\bibitem{14} C. N. Yang, {\it Phys. Rev. Lett}. {\bf 19}(1967)1312;
R. J. Baxter, {\it Ann}. {\it Phys}. {\bf 70}(1972)193.
\bibitem{15} Zhong-Qi Ma, {\it Yang-Baxter Equation and Quantum Enveloping
algebras}, World Scientific, Singapore, 1993.
\bibitem{16} L. D. Faddeev, N. Y. Reshetikhin and L. A. Takhtajan,
Quantization of Lie groups and Lie algebras, in {\it Algebraic
analysis}, Academic Press, 129, 1988.
\bibitem{17} J. S. Bell and R. Jackiw, {\it Phys. Rev}. {\bf D6}(1972)477;
R. Jackiw, Topological investigation of quantized gauge theories,
in {\it Anomalies, Geometry, Topology}, P.211, Eds. W. Bardeen and
A. White, World Scientific, Singapore, 1985.
\bibitem{18} G. Gasper and M. Rahman, Basic Hypergeometric Series,
in {\it Encyclopedia of Mathematics and its Applications 35},
Cambridge, 1990; F. H. Jackson, {\it Quart. J. Pure Appl. Math}.
{\bf 41}(1910)193.
\bibitem{19} Bo-Yu Hou and Yao-Zhong Zhang, {\it J. Math. Phys}.
{\bf 28}(1987)1709.

\end{thebibliography}
\end{document}